\documentclass[a4paper,11pt]{article}

\pdfoutput=1

\usepackage{jcappub}

\usepackage{color} 
\definecolor{Blue}{rgb}{0,0,0.9}
\definecolor{Red}{rgb}{0,0,0.9}
\definecolor{Green}{rgb}{0,0,0.9}

\usepackage[T1]{fontenc}
\usepackage{widetext}
\usepackage{subfig}
\usepackage{booktabs}

\title{\boldmath Turnaround radius in $f(R)$ model}

\author[a,b]{Rafael C. C. Lopes,}
\author[a]{Rodrigo Voivodic,}
\author[a]{L. Raul Abramo,}
\author[c]{Laerte Sodr\'{e} Jr}

\affiliation[a]{Instituto de F\'isica, Universidade de S\~ao Paulo, 
05314-970, S\~ao Paulo, Brazil}
\affiliation[b]{Instituto Federal de Educa\c{c}\~{a}o, Ci\^{e}ncia 
e Tecnologia do Maranh\~{a}o,  Santa In\^{e}s, MA 65304-770, Brazil }
\affiliation[c]{Instituto de Astronomia, Geof\'{i}sica e Ci\^{e}ncias Atmosf\'{e}ricas, Universidade de S\~{a}o Paulo, Rua do Mat\~{a}o, 1226, S\~{a}o Paulo, SP 05508-090, Brazil}

\emailAdd{rafaelcastro@ifma.edu.br}
\emailAdd{rodrigo.voivodic@usp.br}
\emailAdd{abramo@fma.if.usp.br}
\emailAdd{laerte.sodre@iag.usp.br}

\abstract{We investigate the turnaround radius in the spherical collapse model, both in General Relativity and in modified gravity, in particular $f(R)$ scenarios. The phases of spherical collapse are marked by the density contrast in the instant of turnaround $\delta_t$, and by the linear density contrast in the moment of collapse, $\delta_c$. We find that the effective mass of the extra scalar degree of freedom which arises in modified gravity models has an impact on $\delta_t$ of up to $\sim10\%$, and that $\delta_c$ can increase by $\sim1.0\%$. We also compute the turnaround radius, $R_t$, which in modified gravity models can increase by up to $\sim 6\%$ at $z \simeq 0$.}

\keywords{$f(R)$ model, Modified Gravity, Turnaround radius, Spherical Collapse} 

\begin{document}
\maketitle
\flushbottom

\section{Introduction}
\label{sec:intro}

Measurements of the redshifts and distances of type Ia supernovae \citep{riess1998,perlmutter1999} led to the conclusion that the universe is undergoing accelerated expansion. 
The standard way to explain this acceleration is by introducing a new energy component, in addition to dark matter, called Dark Energy (DE)\citep{amendola2010dark}. In the simplest and most popular model, $\Lambda$CDM, a cosmological constant ($\Lambda$) plays the role of DE. However, this solution suffers from old-standing problems \citep{weinberg1989cosmological}, which motivates the investigation of other explanations for cosmic acceleration.

One such alternative is to modify General Relativity (GR) \citep{carroll2004cosmic}. In particular, $f(R)$ models replace the Ricci scalar curvature $R$ of the Einstein-Hilbert action by $R+f(R)$. The distinction between $f(R)$ and $\Lambda$CDM has been studied in many contexts, e.g., in the formation and evolution of stars \cite{Capozziello:2011gm}, in clusters of galaxies \cite{PhysRevD.92.044009}, type Ia supernovae \cite{PhysRevD.93.084016}, the matter power spectrum \cite{oyaizu2008nonlinear}, structure formation \cite{brax2012structure} and voids \cite{voivodic2016modelling}.
 
We focus on a particular class of $f(R)$ models proposed by \cite{sawicki2007stability,hu2007models}, the so-called Hu-Sawicki model. 
In this paper we study the Spherical Collapse Model (SCM) in these theories, comparing it to the results in $\Lambda$CDM. The SCM has been studied in modified gravity by \cite{voivodic2016modelling, PhysRevD.79.084013, Pace11082010, PhysRevD.81.063005, schafer2008spherical, PhysRevD.89.023523, PhysRevD.92.083529}, and specifically in $f(R)$ models by \cite{PhysRevD.85.063518, PhysRevD.88.084015, Chakrabarti2014, cembranos2012gravitational}. 
However, whereas those works are concerned with the linear density contrast in the moment of collapse, $\delta_c$, which is key to the halo mass function and bias, 
we are interested in the instant when an overdensity reaches its maximum size -- the turnaround. 
We calculate $\delta_t$, the non linear matter density contrast at turnaround, as well as $\delta_c$,  both for GR and for $f(R)$ models. 
In particular, we point out that $\delta_t$ is more sensitive to modifications of gravity compared with $\delta_c$. 
Namely, $\delta_t$ can change by up to $\sim10\%$ in the large field limit (when the modifications of gravity are strongest) compared with a shift of only $\sim 2\%$ for $\delta_c$. However, it should be cautioned that the exponential dependence of the mass function on $\delta_c$ means that observations which are sensitive to this quantity may be good discriminants of the gravity model as well.

The moment of turnaround is marked by the maximal radius of the spherical region, which is called the turnaround radius $R_t$. Observationally, this means a spherical surface of null radial physical velocity, $\dot{r}=0$.
In practice, the hierarchical nature of structure formation means that turnaround is happening on increasingly large scales, often far from spherical symmetry, and in regions that include smaller collapsed structures (sub-halos). 

Despite the obvious limitations of the SCM, the turnaround radius $R_t$ is a useful tool to test cosmological models \cite{Pavlidou:2014aia, Pavlidou:2013zha, Tanoglidis:2014lea, Tanoglidis:2016lrj}, including modified gravity models such as DGP \cite{Lee:2016bec}. 
However, these works have focused mainly in the maximum turnaround radius -- the size of the surface where the radial {\it acceleration} is null, $\ddot{r}=0$, which corresponds to an upper bound for the value of the turnaround radius. In that context, Capoziello et al. \cite{2018arXiv180501233C} have recently found an expression for the maximum turnaround radius for any $f(R)$ model, including viability conditions for this class of MG.
From an observational perspective, Lee et al. 2015 \cite{lee2015bound} have estimated a turnaround radius for the galaxy group NGC $5353/4$ which is larger than the maximum turnaround radius which is, in principle, allowed by the $\Lambda$CDM model.

The main motivation for this paper is the real possibility of measuring $R_t$ in forming structures, where one could distinguish between $\Lambda$CDM and models of MG. These observations provide local measurements of the strength of gravity, as a function of both time and distance, and in relatively less dense regions, $\delta \approx 10$, where screening mechanisms should act weakly.
 
This paper is organized as follows: in Section \ref{sec:1} we present $f(R)$ theories, in particular the Hu-Sawicki model. Then in Section \ref{sec:2} we describe the SCM, its main features, and a qualitative discussion of how spherical collapse is influenced by modifications of gravity in the limits of small and large field. In Section \ref{sec:3} we examine in detail the dynamics of spherical collapse, including the radial, time and mass dependence. Then, in Section \ref{sec:4}, we estimate the changes in the density at turnaround, $\delta_t$, including the dependence with time and with mass. In Section \ref{sec:5} we summarize the main results of this work: the changes in the turnaround radius in the Hu-Sawicki model, and how they scale with the time of turnaround and with the mass of the structure. Finally, in section \ref{sec:conc} we present our main conclusions. 
We also detail some features of our methods in Appendix \ref{append}, where we describe the construction of the physical initial profile, in Appendix \ref{append2}, where we plot some particular density profiles at several stages during its evolution
and in the Appendix \ref{append3} we presented the evolution of the shell on the turnaround and compared it with the maximum turnaround radius.

\section{Modified gravity and structure formation}
\label{sec:1}
\subsection{\texorpdfstring{$f(R)$}{f(R)} theories}
The simplest way to change GR is by modifying the Einstein-Hilbert action with a function of the Ricci scalar $R$.
There is a class of such models called $f(R)$ theories -- see \cite{sotiriou2010f} for a review -- whose action is  written as:
\begin{equation}
S =  \int{d^4 x \sqrt{-g} \left[ \frac{R+f(R)}{16\pi G} + \mathcal{L}_{m} \right]} \, ,\label{eqn:action}
\end{equation}
where $\mathcal{L}_{m}$ is the matter Lagrangian density. 
The modified Einstein equations are:
\begin{equation}
G_{\alpha \beta} + f_R R_{\alpha \beta} - \left(\frac{f}{2} - \Box f_R \right)g_{\alpha \beta}- \nabla_{\alpha} \nabla_{\beta} f_R = 8\pi G \, T_{\alpha \beta} \, , \label{egm}
\end{equation}
where $f_{R} \equiv \frac{d f(R)}{dR}$, $G_{\alpha \beta}$ is the Einstein tensor, and $T_{\alpha \beta}$ is the energy-momentum tensor. If we replace $f(R) \to -2\Lambda$, the $\Lambda$CDM model is recovered.

From the trace of \eqref{egm} we obtain a Klein-Gordon equation for $f_R$, which can be regarded as a scalar field. This field is coupled to the metric in the form:
\begin{equation}
3 \Box f_{R} - R + f_{R}R - 2f = -8\pi G \, T \, , \label{eqfield}
\end{equation}
where $T$ is the trace of the energy-momentum tensor. The effective potential \citep{sotiriou2010f} is defined as:
\begin{equation}
{\frac{\partial V_{\rm eff}}{\partial f_R}} \equiv \Box f_{R}= {\frac{1}{3}}\left( R- f_R R + 2 f - 8\pi G \, T\right) \, , \label{potn}
\end{equation}
and the mass of the scalar field is:
\begin{equation}
m^2_{f_R}\equiv{\frac{\partial^2 V_{\rm eff} }{\partial f_R^2}}=\frac{1}{3}\left[\frac{1+f_{R}}{f_{RR}}-R\right] \, .
\label{eq:massadocampofr}
\end{equation}
This expression can be simplified if we impose the condition that $|Rf_{RR}|\cong f_R <<1$, which is necessary for mimic the $\Lambda$CDM background expansion history.
\begin{equation}
m^2_{f_R}\approx  \frac{1}{3 f_{RR}} \, .
\label{eq:massadocampescalaprox}
\end{equation}

The inverse of this mass scale defines the comoving Compton wavelength $\lambda_c \equiv \frac{1}{m_{f_R}}$ that describes the range of the scalar field interactions. 
Since the scalar field, according to Eq. \eqref{potn}, couples to all forms of matter whose energy-momentum tensors have non-zero traces, we can regard it as giving rise to a fifth force, in addition to the gravitational force described by GR.
When the scalar mass is small, the reach of the fifth force is longer, and conversely, if the mass is large (e.g., in regions of high density), $\lambda_c$ is small and the range of the fifth force is more limited. 
The mechanism whereby the scalar field acquires a mass in dense regions such as stars and planets is known as the chameleon mechanism \cite{khoury2004chameleon}, and for this reason these modified gravity models can evade solar-system tests of GR. 

Typically we are interested in matter configurations which can present large spatial gradients, but that are far from the relativistic regime -- i.e., the spatial gradients dominate over time derivatives. 
In this ``quasistatic'' regime, after subtracting the background, Eq. \eqref{eqfield} becomes:
\begin{equation}
\nabla^2 \delta f_{R}=\frac{a^2}{3}\left[\delta R  -8\pi G \delta \rho_m  \right] \, ,
\label{eq:eqpoissontofr}
\end{equation}
where $\delta f_{R} = f_{R}(R)-f_{R}(\bar{R})$, $\delta R = R-\bar{R}$, and $\delta \rho_m = \rho_m-\bar{\rho}_m$ (bars represent spatial averages). 
Furthermore, when we consider the time-time component of Eq. \eqref{egm}, in the Newtonian gauge, we obtain the modified Poisson equation in comoving coordinates:
\begin{equation}
\nabla^2 \Phi = \frac{16 \pi G}{3}a^{2} \delta \rho_{\rm m} - \frac{a^{2}}{6} \delta R(f_R) \, . \label{potorig}
\end{equation}
The previous equations, combined with the relativistic fluid equations for the matter density fields, form a closed system.

If the perturbation of the scalar field is small, then a linear approximation leads to $\delta R \approx (dR/df_R)\delta f_R $. [Notice that the chameleon mechanism cannot be applied within the linear approximation for the scalar field.] In Fourier space we have, then:
\begin{equation}
- k^2 \Phi = \left[1+ \epsilon(a,k) \right] 4 \pi G a^2 \delta \rho_m \, , 
\label{eq:eqpoissonfrconjM}
\end{equation}
where:
\begin{equation}
\epsilon(a,k) \equiv \frac{1}{3} \left(\frac{k^2}{a^2 m_{f_R}^2 + k^2}\right) \, .
\label{eq:defepsilonM}
\end{equation}
For this reason, a common approach to represent the effects of the $f(R)$ model is to describe it as a modification of Newton's constant $G$, which, in the linear approximation and in Fourier space, is written as $G_{eff} \equiv G \left[1+ \epsilon(a,k)\right]$. The term $\epsilon(a,k)$ in Eq. \eqref{eq:defepsilonM} describes the modifications of Einstein's gravity, in such a way that, when it vanishes, we recover GR. 
This limit can also be reached when the scale of interest, 
$\lambda$, is larger than the Compton wavelength $\lambda_c$, i.e.,  or equivalently $a^2 m_{f_R}^2 >> k^2$. 
This is the so-called small-field (SF) limit. 
On the other hand, if $\lambda << \lambda_c$ ($a^2 m_{f_R}^2 << k^2$), the effects of the fifth force are maximal, and $\epsilon(a,k)=1/3$. This is known as the large-field (LF) limit.

\subsection{Hu-Sawicki model}
The Hu-Sawicki model \cite{hu2007models, hu2007parametrized} can be written as:
\begin{equation}
f(R) = -m^2 \frac{c_{1}(R/m^2)^n}{c_{2}(R/m^2)^n + 1} \, ,
\label{frhusawick0}
\end{equation}
where $c_1$, $c_2$ and $n$ are dimensionless free parameters and $m^2 \equiv 8 \pi G \bar{\rho}_{\rm M}/{3}$. 
In the high curvature regime, expanding Eq. \eqref{frhusawick0} and taking $n=1$, we obtain:
\begin{equation}
f(R) \approx -16 \pi G \rho_\Lambda - f_{R0} \frac{\bar R_0^2}{ R} \, ,
\label{fRapprox}
\end{equation}
where the dominant constant term has been matched to a cosmological constant, 
$\bar{R}_0=\bar{R}(z=0)$ and $\bar{f}_{R0}=f_R(\bar{R}_0)$. 
In this way, we recover the $\Lambda$CDM model if $|\bar{f}_{R0}|=0$.
It is also useful to rewrite the mass of the scalar field in Eq. \eqref{eq:massadocampescalaprox} as:
\begin{equation}
m_{f_R}\approx \frac{H_0}{c} \left(\frac{\Omega_{m0}  + 4\Omega_\Lambda}{2|f_{R0}|}\right)^{1/2} \left(\frac{\Omega_{m0} (1+z)^3 + 4\Omega_\Lambda}{\Omega_{m0}  + 4\Omega_\Lambda}\right)^{3/2} \, .
\label{eq:paramMfr}
\end{equation}

In the following we will study the evolution of matter perturbations, as described by the equations of the previous Section, in the context of the Hu-Sawicki model. 
When necessary, we assume a cosmological model with parameters $\Omega_m=0.31$, $\Omega_\Lambda=0.69$, $\sigma_8=0.86$ and $n_s=0.96$.

\section{Spherical collapse in \texorpdfstring{$f(R)$}{f(R)}}
\label{sec:2}
Starting from the Euler equation together with the continuity equation for a nonrelativistic pressureless fluid, and assuming an initial spherically symmetric density profile, the nonlinear equation that describes the evolution of the density contrast is:
\begin{equation}
\delta''+\left( {3\over a}+{H'\over H} \right)\delta'-{4 \, {\delta'^{2}}\over {3(1+\delta)}}=\frac{1+\delta}{H^2 \, a^4} \nabla^2 \Phi 
\, ,
\label{deltanonlinear}
\end{equation}
where ${}'= d/da$. 
Following the prescription of \citep{brax2012structure} to express the potential in configuration space, Eq. \eqref{eq:eqpoissonfrconjM} can be rewritten as:
\begin{equation}
\Phi(\vec{x},a)=-4 \pi G a^2 \bar{\rho}_m \int \frac{d^3k}{(2 \pi)^3}
e^{i\vec{k}\cdot\vec{x}} \, k^{-2} \, [1+\epsilon(k,a)] \delta(\vec{k},a) \,.
\label{potentialphix}
\end{equation}
Under the assumption of spherical symmetry, and using $H(a) = H_0 E(a)$, Eq. \ref{deltanonlinear} can then be recast as:
\begin{equation}
\delta''+\left({3\over a}+{E'\over E}\right)\delta'-{4{\delta'^{2}}\over {3 (1+\delta)}}=\frac{3(1+\delta)}{2 \, E^2 \,2 \pi^2}  \,\Omega_{m0}\,a^{-5}\int_0^{\infty} dk \, k^2 [1+\epsilon(k,a)] \delta(k,a) \frac{\sin(kr)}{k r} \,.
\label{deltanonlinearx}
\end{equation}
By solving this equation, as well as its linearized version, we are able to determine all relevant properties of spherical collapse in $f(R)$ modified gravity. 

\subsection{Dynamics of the matter density contrast}
\label{solucofdelta}
In order to compute the density contrast $\delta(r)$ it is necessary to consider, firstly, an initial density profile $\delta_i(r)$, which will be evolved from an early time, where we chose $z_i = 500$, because in this time the MG is not distinguishable from GR,  until the collapse time. The source term of the right-hand side of Eq. \eqref{deltanonlinearx} includes the gravitational interactions, including the modified gravity corrections through $\epsilon(k,a)$. We start with the profile in Fourier space at each time step, compute the corrections, recalculate the Fourier space integral, and so on, until the turnaround -- or until collapse.

Following \citep{PhysRevD.88.084015}, we implemented two initial profiles. The first was a generic initial density profile (Tanh), characterized by the size of the top-hat-like function $r_b$ and the steepness of the transition $s$, such that in the limit  $s\rightarrow 0$ we recover the top-hat profile:
\begin{equation}
\delta_i(r)=\frac{\delta_{i,0}}{2}\left[1-\tanh\left(\frac{r/r_b-1}{s}\right)\right].
\end{equation}
The mass inside the profile is well approximated by $M_b=(4\pi/3)\bar{\rho}_0 r_b^3$, since the extra contribution of $\delta(r)$ to the mean initial density of the spherical region is negligible at the initial surface.

The second profile we consider is a physical mean density profile (Phy) around a Gaussian density peak of height $\nu\equiv\delta_{i,0}/\sigma_i(R)$:
\begin{equation}
\delta_i(r,R)=\left\langle \delta(z_i,\textbf{x},R)|peak,\nu \right\rangle
\label{defphydeltai}
\end{equation}
where $\delta(z_i,\textbf{x},R)$ is a Gaussian random field smoothed with a top-hat window function $W(kR)$ inside a comoving scale $R$ -- for details, see \cite{bardeen1986statistics}. The initial density profile can then be written as:
\begin{equation}
\delta_i(r,R)=\frac{2}{\pi} \int_0^\infty d k k^2 \delta_0(k,R) \frac{\sin k r}{k r} T(k)\,
\label{fourieinverofprofile}
\end{equation}
where $T(k)$ is the matter transfer function and $\delta_0(k,R)$ is the primordial shape function in $k$ space, which is given by \citep{PhysRevD.88.084015}:
\begin{equation}
\delta_0(k,R)=\delta_{i,0}\frac{1}{4} \pi  (n_s+5) R^3 e^{-k^2 R^2} (k R)^{n_s}\cdot F(\nu,n_s,k,R)  \, ,
   \label{peaksshapeprimordial1}
\end{equation}
where $n_s$ is the scalar spectral index and the expression for the function $F(\nu,n_s,k,R)$ is presented in Appendix \ref{append}.
For details about the construction of this initial density profile, see Appendix C of Ref. \cite{PhysRevD.88.084015}. It is important to note that this profile is the mean of many density profiles around peaks of the density field, obtained from a Gaussian realization. We also study how the dynamics of different spherical profiles change by varying the shape parameter $s$ of the Tanh profile.

\subsection{Properties of Spherical Collapse}
\label{propertcollapsespher}
Since the mass inside the spherical region $M=4\pi \rho\, r^3/3$ is constant, the radius of that region is related to the density contrast by $r^3 \bar\rho (1+\delta) = 3M/4\pi$. 
The moment of turnaround is defined by the condition that $\left. \dot{r} \right|_{a_t}=0$, which leads to an equation for the scale factor at that instant:
\begin{equation}
\frac{3}{a_t}\left(\delta_t+1\right)=\delta'_t. \label{eq:momentoturnaround}
\end{equation}
With $a_t$, we obtain the density contrast in the moment of turnaround, $\delta_t=\delta(a_t)$. 

At a later time that structure collapses, at an instant $a=a_c$, when  the linearized density contrast takes its ``critical'' value, $\delta_c=\delta_{l}(a_c)$. Here we consider that the collapse happens when the non-linear density contrast reaches the value $\delta_c=200$ -- see \citep{valageas2009mass}, because this value choice has a better agreement with N-body simulations in the halo mass function  \citep{valageas2009mass}.

\subsection{Small- and large-field limit}
We now describe how the modified gravitational force, which can increase by a factor of up to $1/3$, impact the parameters $\delta_c$ and $\delta_t$. 
We do this both in the context of the small-field (SF) limit, which represents unmodified gravity and is characterized by $\epsilon \to 0$ (GR), and in the large-field (LF) limit, which represents the maximal effects of modified gravity, and is characterized by $\epsilon \to 1/3$, in the context of the linear approximation. Those quantities (which in these two limits are independent of scale and mass) were computed in the case of structures that collapse and reach the turnaround at $z=0$ ($a=1$), in the two limits. 

The results are shown in table \ref{tab1} for both profiles, Tanh and Phy.
$\Delta_{rel}$  shows the fractional differences between the values in each case relative to the values with $\epsilon=0$.
Notice that the density contrast at the moment of collapse $\delta_c$ increases by $\sim 1.0\%$. 
On the other hand, the density contrast at the moment of turnaround, $\delta_t$, decreases by $\sim 20\%$, in both profiles.

\begin{table}[ht]
\centering
\begin{tabular}{ccccc}
\hline
\textbf{Tanh(s=0.4)/Phy} & \textbf{$a_c$} & \textbf{$a_t$} & \textbf{$\delta_t$} & \textbf{$\delta_{c}$} \\
\hline
$\epsilon=0$ & 1.0 & 0.609/0.609 & 6.1025/6.1104 & 1.598/1.598 \\
\hline
 $\epsilon=1/3$ & 1.0 & 0.614/0.613 & 4.930/4.928 & 1.609/1.609 \\
 \hline
$\Delta_{rel}$ & --  & --  & 0.192/0.194 & 0.0069/0.0069 \\
\hline
 &  &  &  &  \\
\hline
$\epsilon=0$ & 1.810/1.811 & 1.0 & 10.912/10.912 & 1.586/1.586 \\
\hline
 $\epsilon=1/3$ & 1.822/1.824 & 1.0 & 8.641/8.647 & 1.597/1.598 \\
 \hline
$\Delta_{rel}$ & --  & --  & 0.208/0.208 & 0.0068/0.0076 \\
\hline
\end{tabular}
\caption{\small{Values of $\delta_{c}$ and $\delta_t$ in the small-field limit $\epsilon=0$ and in the large-field limit $\epsilon=1/3$, for structures of $M\approx 10^{14} \, h^{-1} M_\odot$, with turnaround and collapse at $a=1$ ($z=0$), for the Tanh profile with $s=0.4$, and for the Phy profile. $\Delta_{rel}$ is the relative difference between the quantities in the large-field and in the small-field limits.}} 
\label{tab1}
\end{table}

\begin{figure}[ht]
\centering
\includegraphics[width=3.0 in]{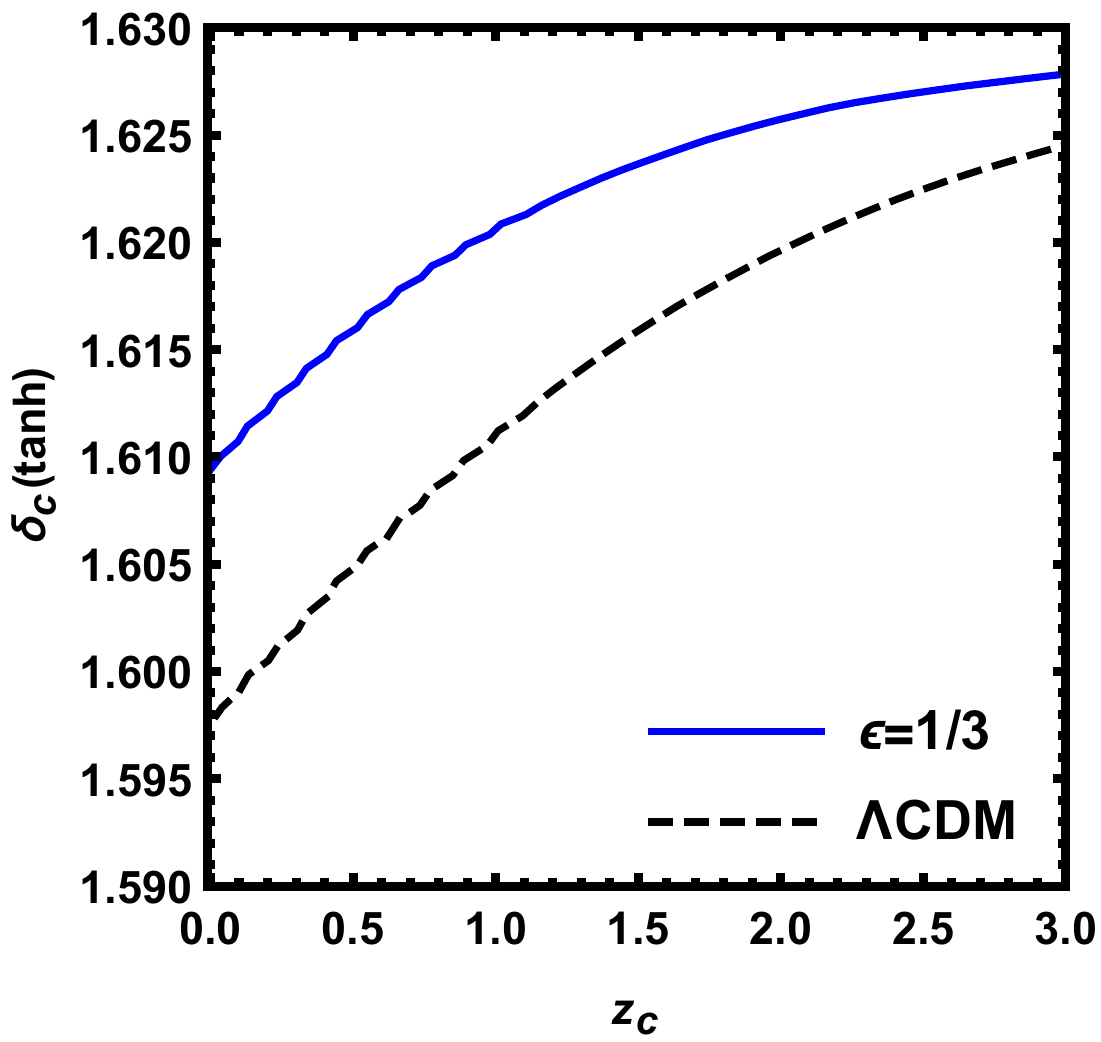}
\includegraphics[width=3.0 in]{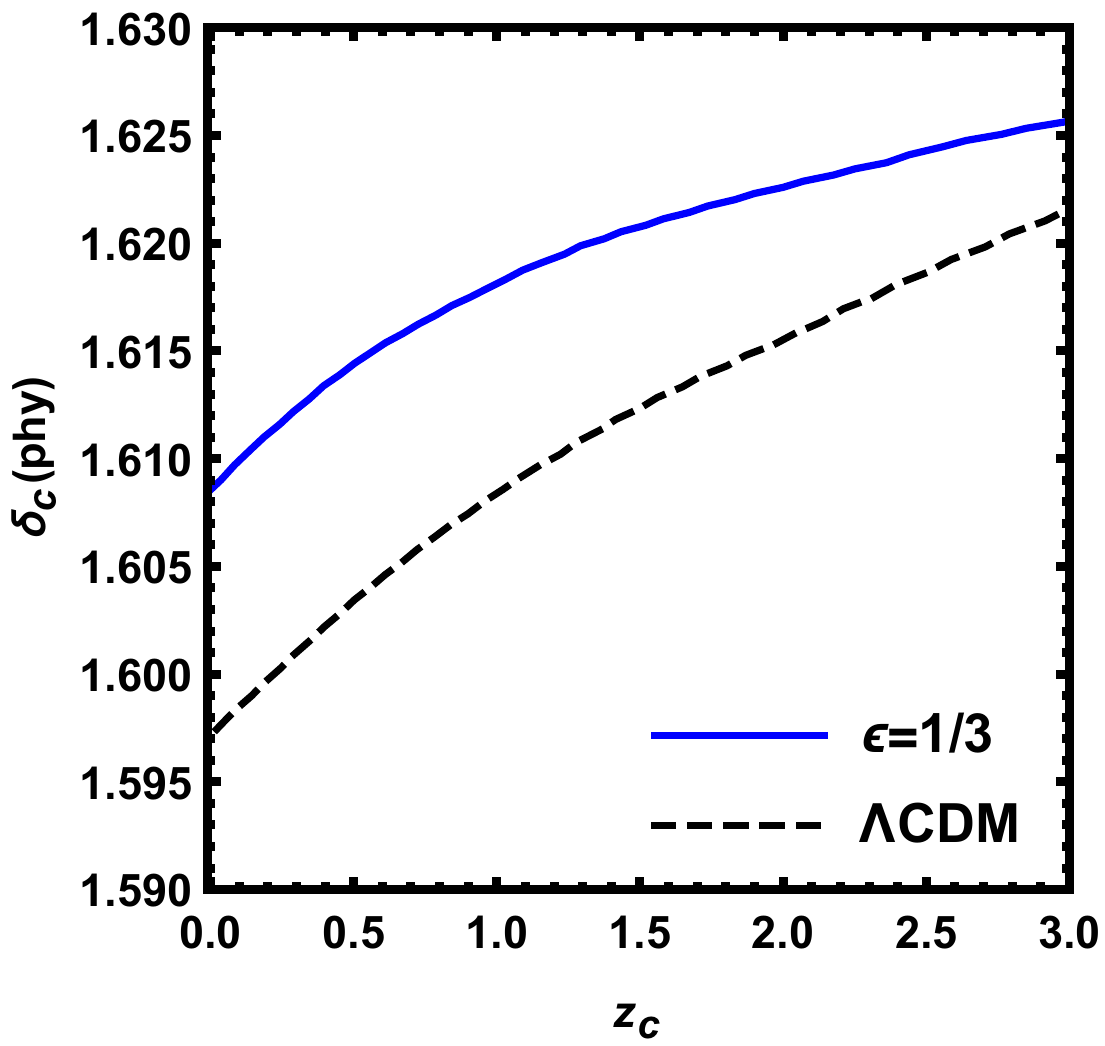}
\includegraphics[width=3.0 in]{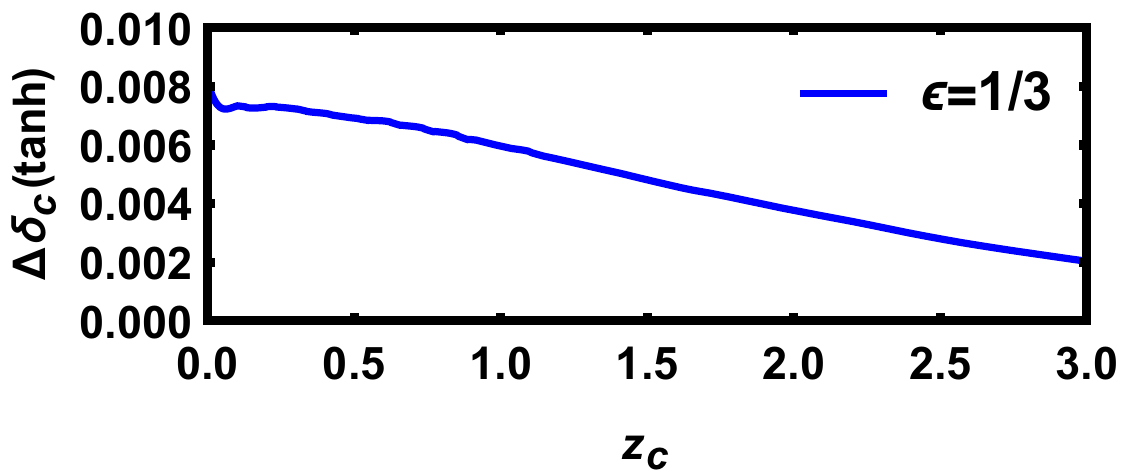}
\includegraphics[width=3.0 in]{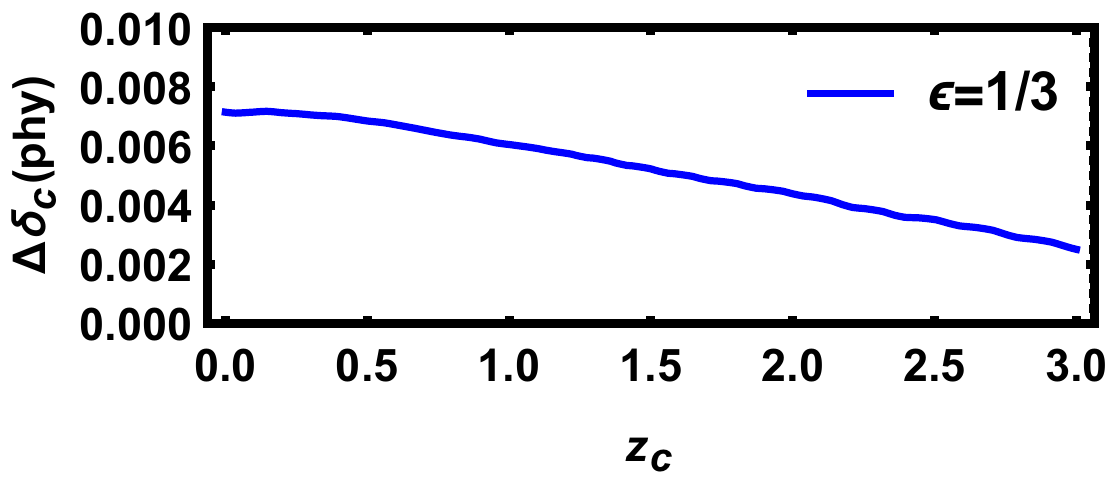}
\caption{Critical density $\delta_c$ of structures with
mass $ M = 10^{14} \, h^{-1} M_\odot$, which collapse in different moments, in the small-field limit ($\epsilon=0$, black-dashed) and in the large-field limit ($\epsilon=1/3$, blue), for the Tanh (left) and Phy (right) profiles. In the bottom we show the relative difference between the values on each limit.}
\label{fig:compadeltace0e13}
\end{figure}

\begin{figure}[ht]
\centering
\includegraphics[width=3.0 in]{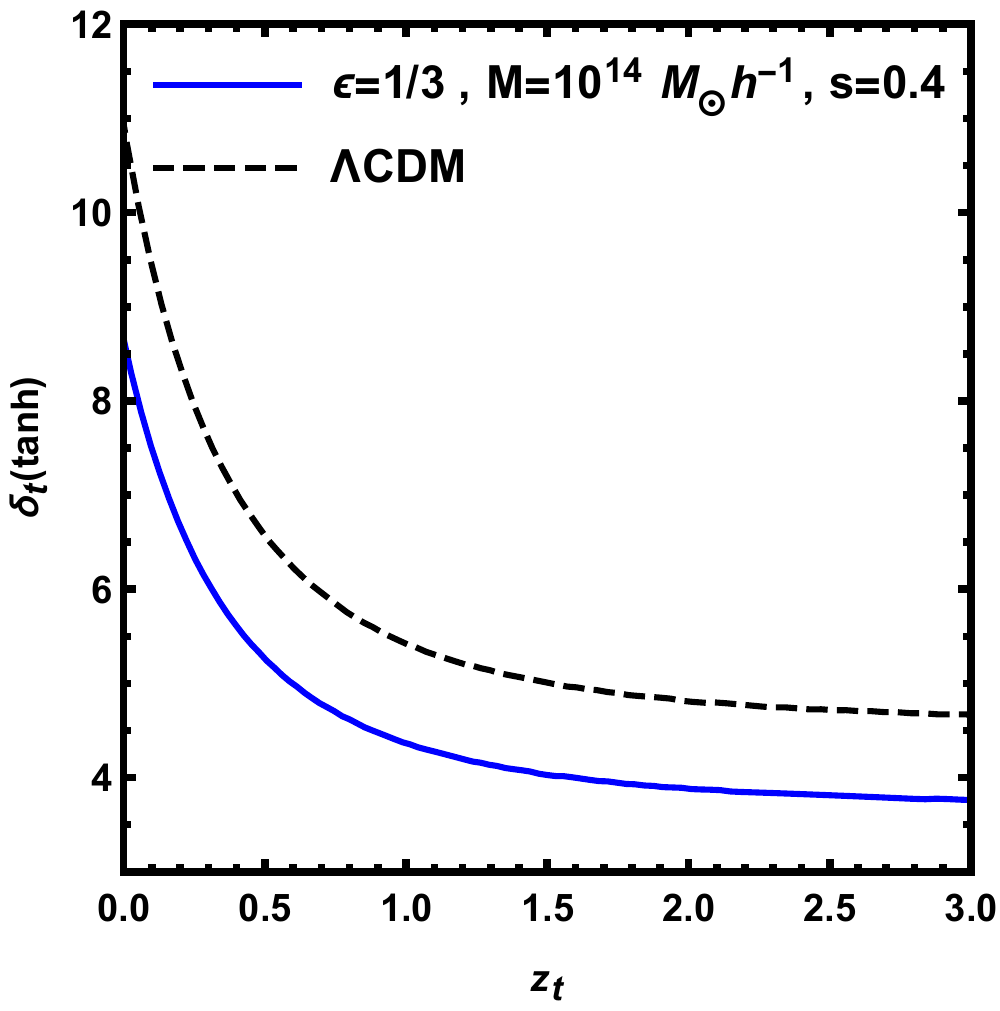}
\includegraphics[width=3.0 in]{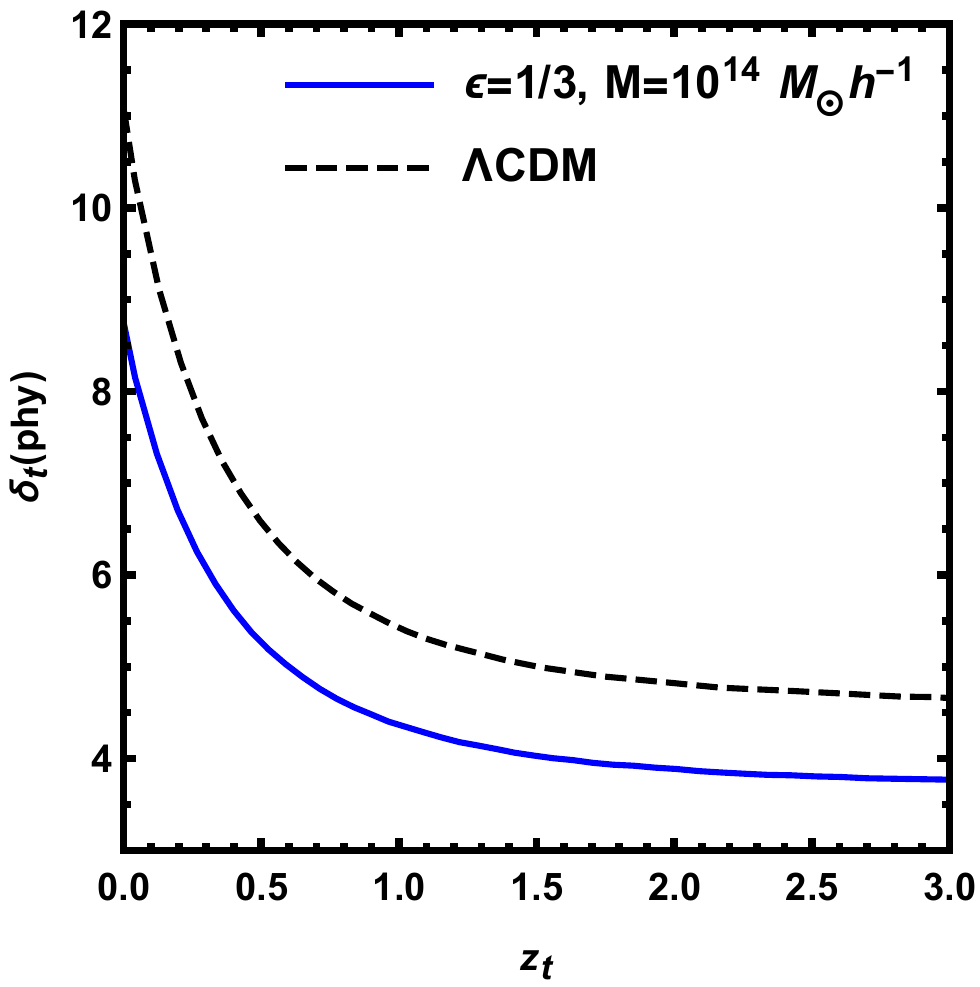}
\includegraphics[width=3.0 in]{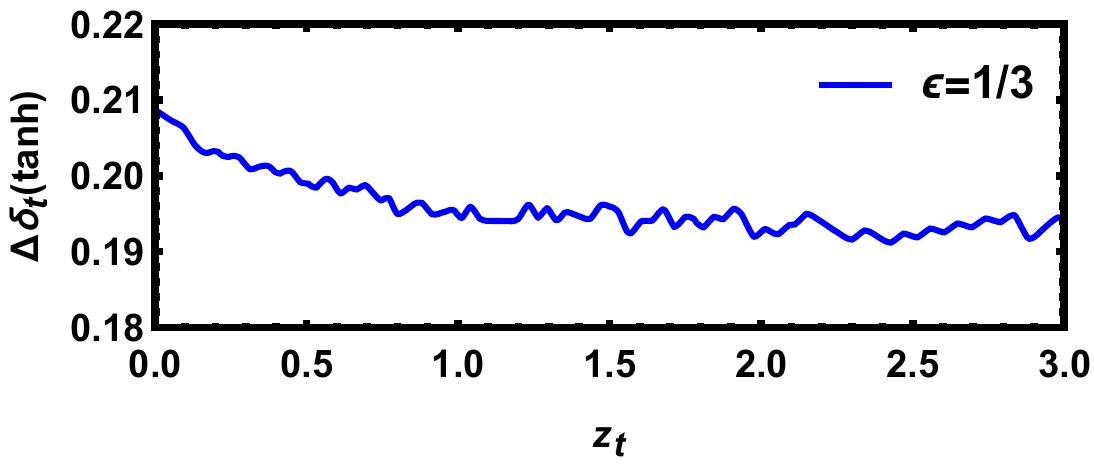}
\includegraphics[width=3.0 in]{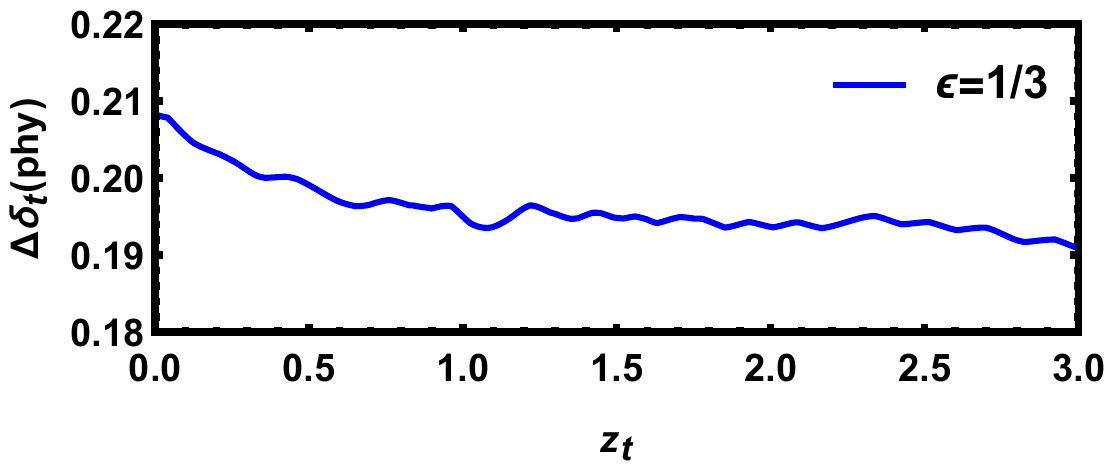}
\caption{Density at turnaround, $\delta_t$, for structures 
with masses $ M = 10^{14} \, h^{-1} M_\odot$,
which reach turnaround at different moments. Captions are the same as in Fig. {\ref{fig:compadeltace0e13}}.}
\label{fig:compadeltate0e13}
\end{figure}

Although Table \ref{tab1} refers to structures which collapse or reach turnaround at $z=0$, it is also interesting to study how spherical collapse takes place at different times in modified gravity.
This is shown in figures \ref{fig:compadeltace0e13} and \ref{fig:compadeltate0e13} (the wiggles seen in these and subsequent plots arise from the discretization of our numerical integration, but they do not affect the mean trends seen in those plots).
It can be seen from Fig. \ref{fig:compadeltace0e13} that $\delta_c$ is larger (by up to $\sim 1.0\%$ -- see the lower panel) when $\epsilon=1/3$ at $z_c=0$. This follows 
from the enhancement of the gravitational force, which aggregates matter more effectively. 
The results for both profiles are very similar, as they should be, since the birkhoff theorem is valid in the two limiting cases and any difference observed is due to numerical effects.

On the other hand, the density contrast at moment of turnaround, $\delta_t$, decreases in the LF limit -- see the left plot of Fig. \ref{fig:compadeltate0e13}.
This can be understood as a result of the increased gravitational force, which makes the turnaround happen sooner compared with the situation when gravity is weaker.
The bottom chart shows that the maximum relative difference is $\sim21\%$, which occurs for structures whose turnarounds happen around $ z\sim0$. 

\section{Effects of \texorpdfstring{$f(R)$}{f(R)} on collapse}
\label{sec:3} 
Until now we have been considering only the SF and the LF limits, characterized by $\epsilon=0$ and $\epsilon=1/3$, respectively.
However, the effects of MG on the spherical collapse at different scales change as a function of time through $\epsilon=\epsilon(k,a)$.

When we include the scale- and time-dependent MG term, it is necessary to solve the equation \eqref{deltanonlinearx} by computing the Fourier integral of the right-hand term at each step in time. In other words, each scale will be affected by all other scales. 
Our approach was based on the method implemented by \cite{brax2012structure} in the context of the spherical collapse model in MG using the initial profile suggested by \cite{PhysRevD.88.084015}.
Therefore, the dependence of $\delta_c$ and $\delta_t$ with the parameter $|f_{R0}|$, which describes the intensity of scalar field $f_R$ today, and therefore of the gravitational field, must be explored together with its mass dependence -- especially since Birkhoff's theorem does not apply in the context of modified gravity.

For $|f_{R0}|$, we have chosen putative values of $10^{-4}$, $10^{-5}$ and $10^{-6}$, since they cover the usual
range of values which are tested in MG models.
We also explore the range of masses $10^{11} \, M_{\odot}\,h^{-1} < M < 10^{16} \, M_{\odot}\,h^{-1} \,$, which includes structures of interest in the nearby Universe.
It is important to note that in this range we could expect effects due to the chameleon mechanism, which is not considered here as the overdensities at the turnaround radius are typically of order one.

\subsection{Time dependence}
\begin{figure}[ht]
\centering
\includegraphics[width=3.0in]{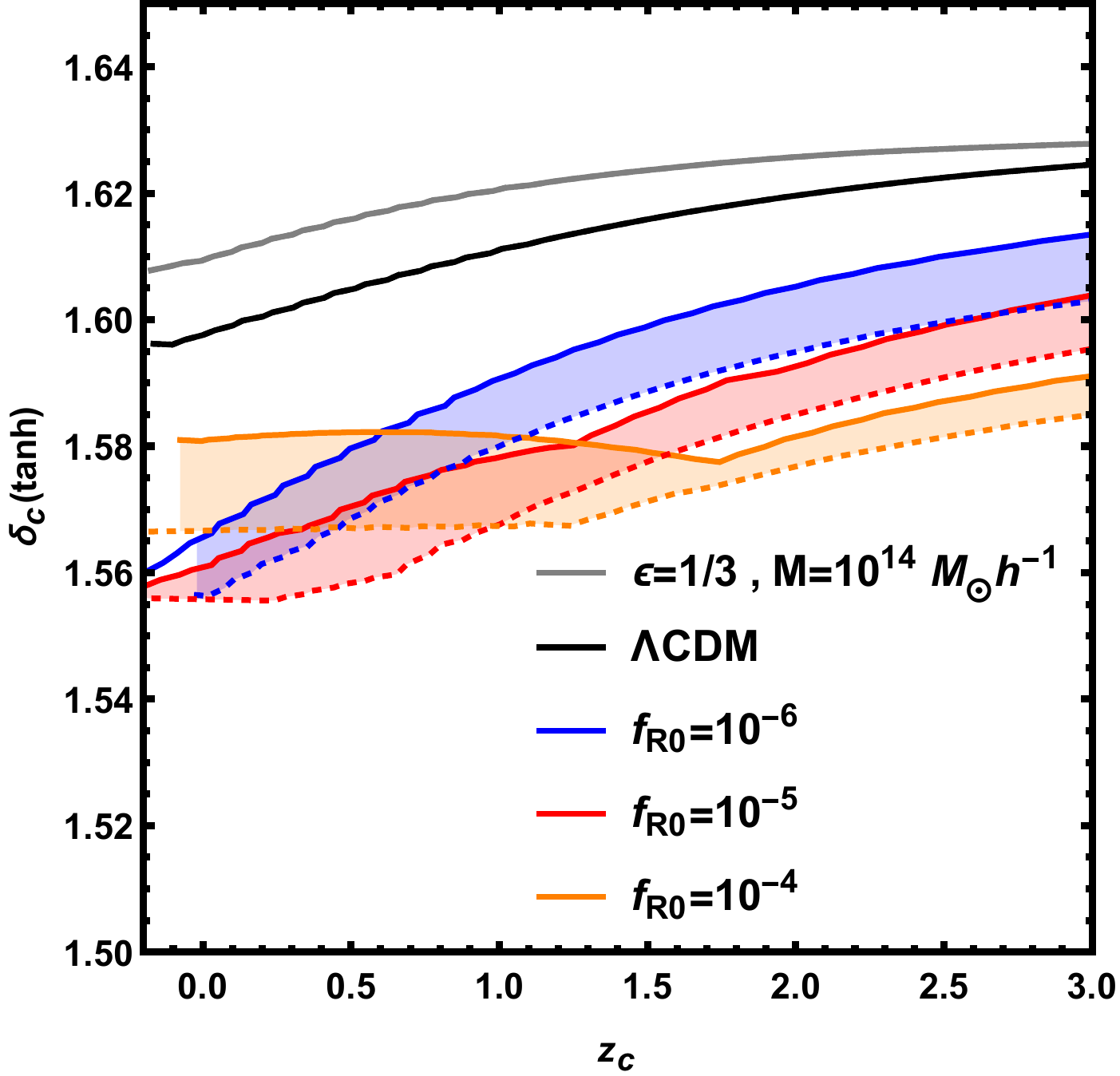}
\includegraphics[width=3.0in]{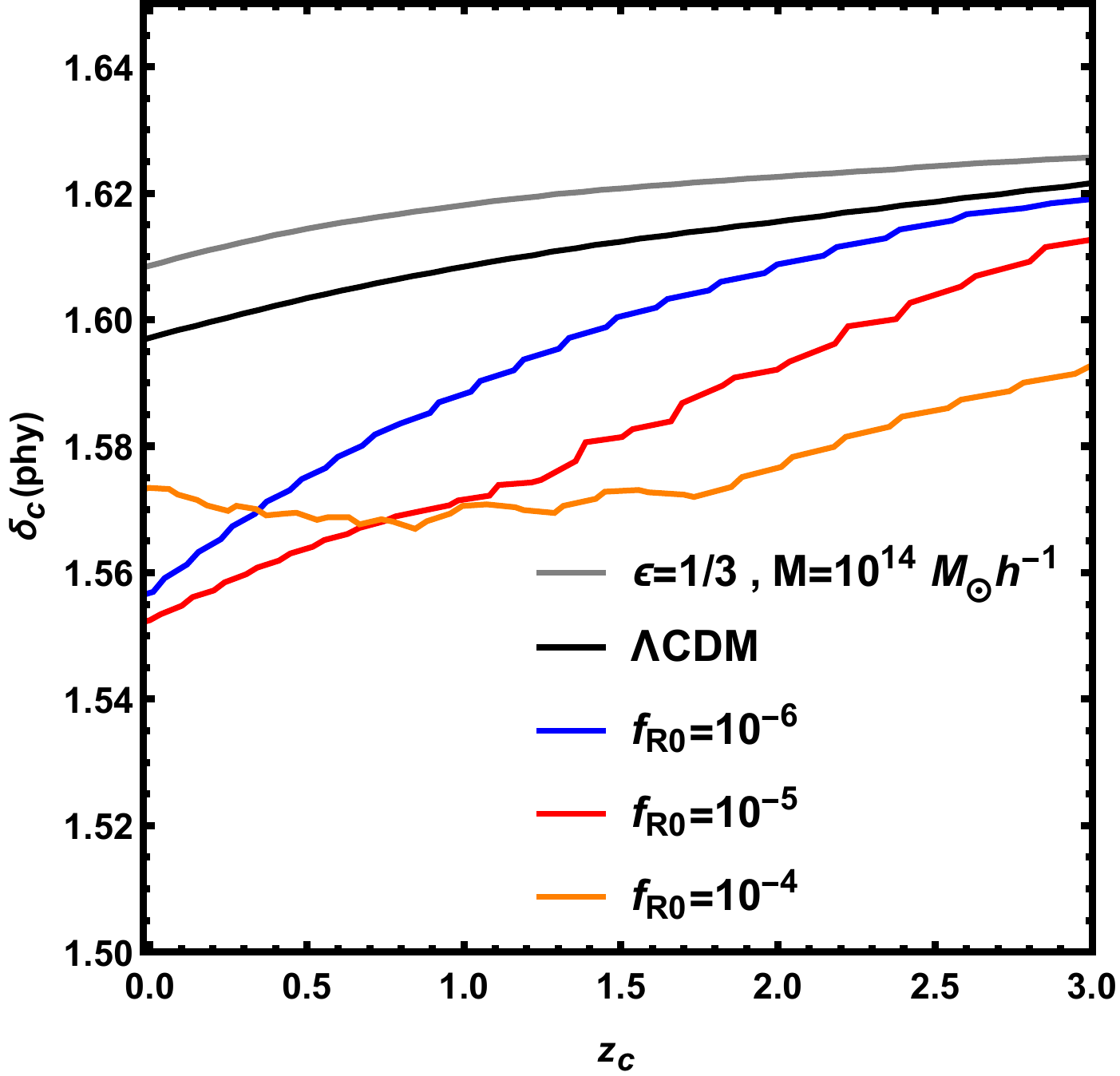}
\includegraphics[width=3.0in]{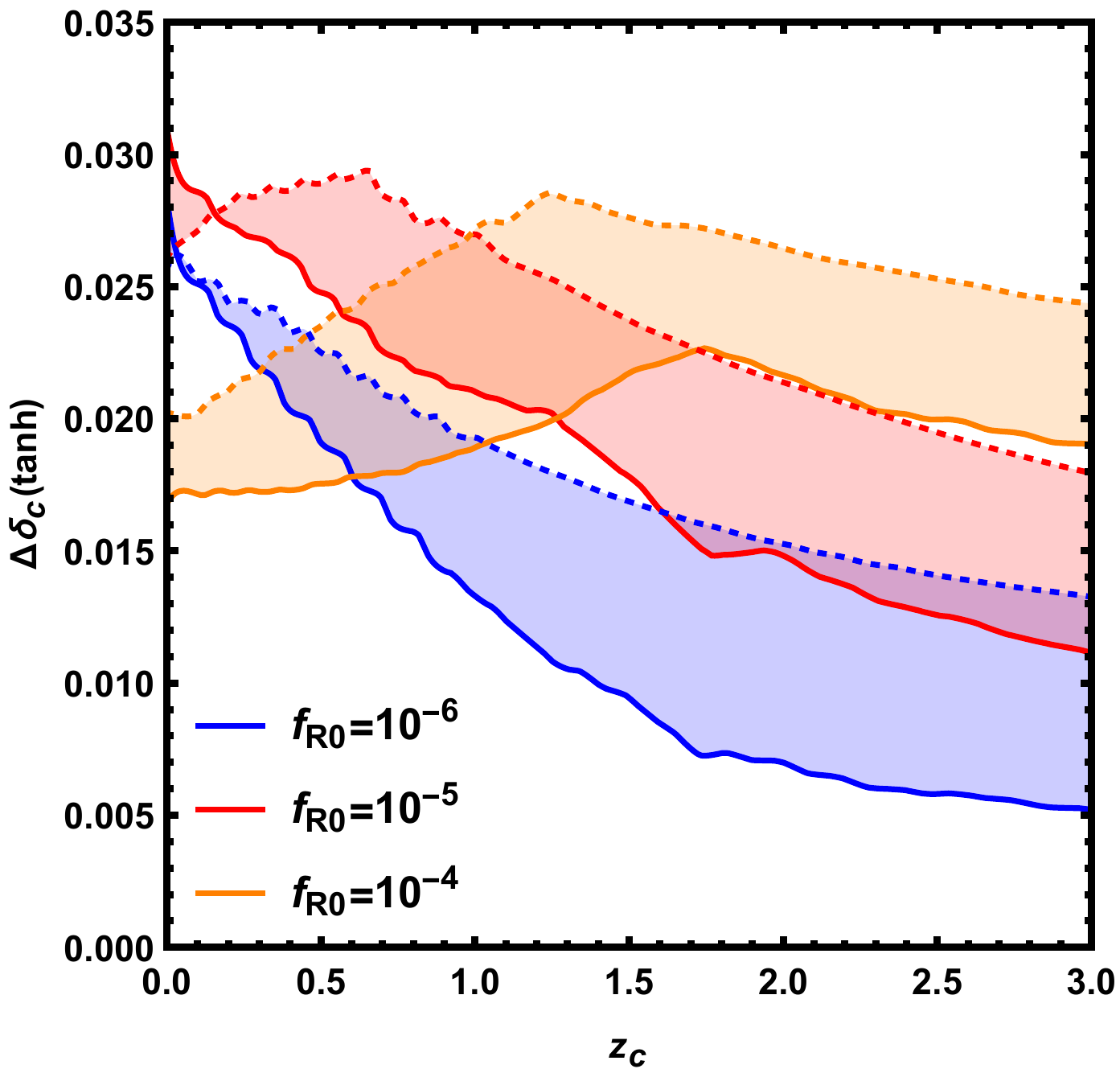}
\includegraphics[width=3.0in]{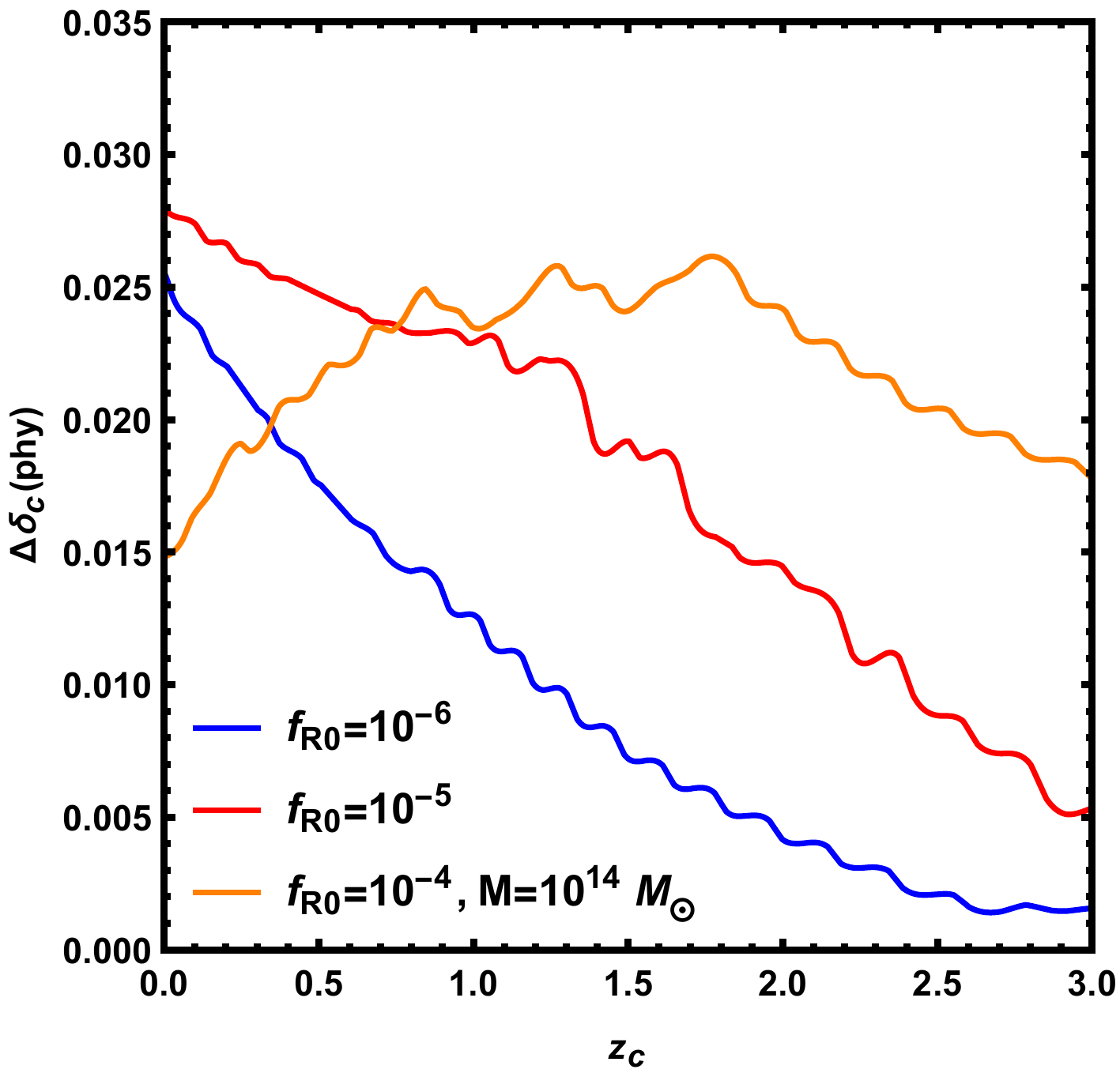}
\caption{\small{Top panels, from left to right: values of $\delta_c$ for the Tanh and Phy profiles, respectively, for $0<z<3$, and for MG parameters $|f_{R0}|=10^{-4}$ (orange), $|f_{R0}|=10^{-5}$ (red), $|f_{R0}|=10^{-6}$ (blue) $\Lambda$CDM, $\epsilon=0$ (Black) and $\epsilon=1/3$ (Gray). The Tanh profile with a smooth slope ($s=0.8$) is plotted as the dotted lines, and the hard slope ($s=0.4$) is plotted as the solid lines.
Lower panels: relative differences between the values of the top panels with respect to $\Lambda$CDM.}}
\label{fig:compadcZdifRel}
\end{figure}

Fig. \ref{fig:compadcZdifRel} shows $\delta_c$ for several MG parameters: $|f_{R0}|=10^{-4}$ (orange), $|f_{R0}|=10^{-5}$ (red), $|f_{R0}|=10^{-6}$ (blue), $\epsilon=0$ (black) and $\epsilon=1/3$ (gray). In the case of the Tanh profile (left panels) we also study how the shape of the profile affects the turnaround and collapse, by employing two different slopes: $s=0.8$ (dotted) and $s=0.4$ (solid). 

An interesting effect appears here, which was first seen by \cite{brax2012structure, PhysRevD.88.084015, PhysRevD.85.063518, voivodic2016modelling}: the values of the density contrast at collapse are below the SF limit. This shows that, by considering the time- and scale-dependence of MG, we are able to reveal effects which do not appear in the simplified cases of the LF and SF limits.
Moreover, the relative difference of $\delta_c$ in MG with respect to  $\Lambda$CDM reaches a maximum of $\sim 3.0\%$ for $|f_{R0}|=10^{-5}$ with $s=0.4$, while for $|f_{R0}|=10^{-6}$ it does not exceed $\sim 2.7\%$. 
We also note that the influence of the slope is largest at $z\sim0$ for $|f_{R0}|=10^{-4}$, reaching $\sim 2.0\%$ for $s=0.8$, and smallest for $|f_{R0}|=10^{-6}$, when the relative change is at most $\sim 2.5\%$ with $s=0.8$.

Another manifestation of the changing effects of MG is the fact that sometimes the curves in Fig. \ref{fig:compadcZdifRel} cross each other. This can be understood as a result of the dynamical relationship between the Compton wavelength of the scalar field, $\lambda_c$, which depends both on $f_{R0}$ as well as on time, and the size of the perturbation. The net, total effect on $\delta_c$ depends on the balance between these length scales integrated over time.   

\subsection{Mass dependence}
\begin{figure}[ht]
\centering
\includegraphics[width=3.0in]{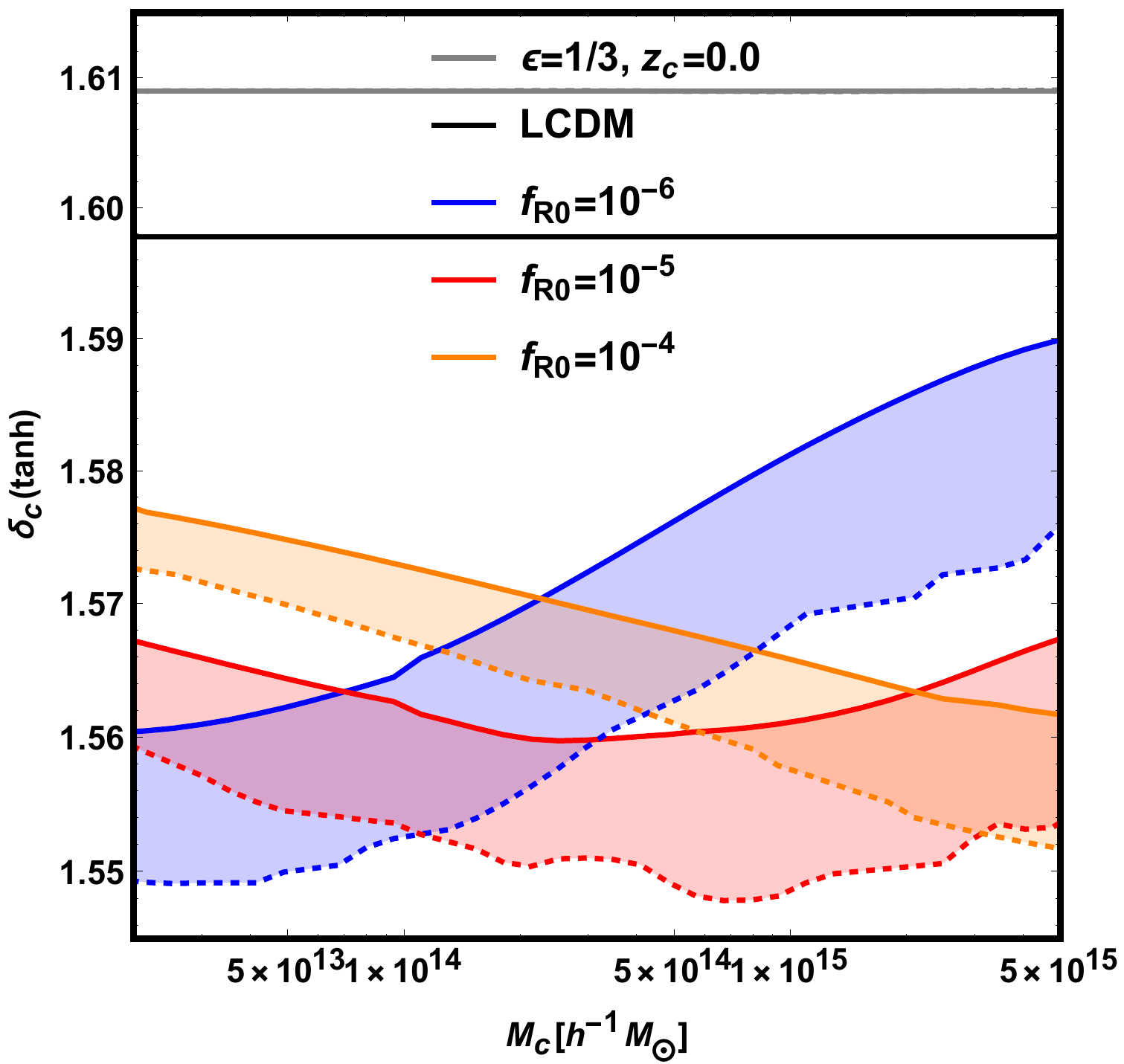}
\includegraphics[width=3.0in]{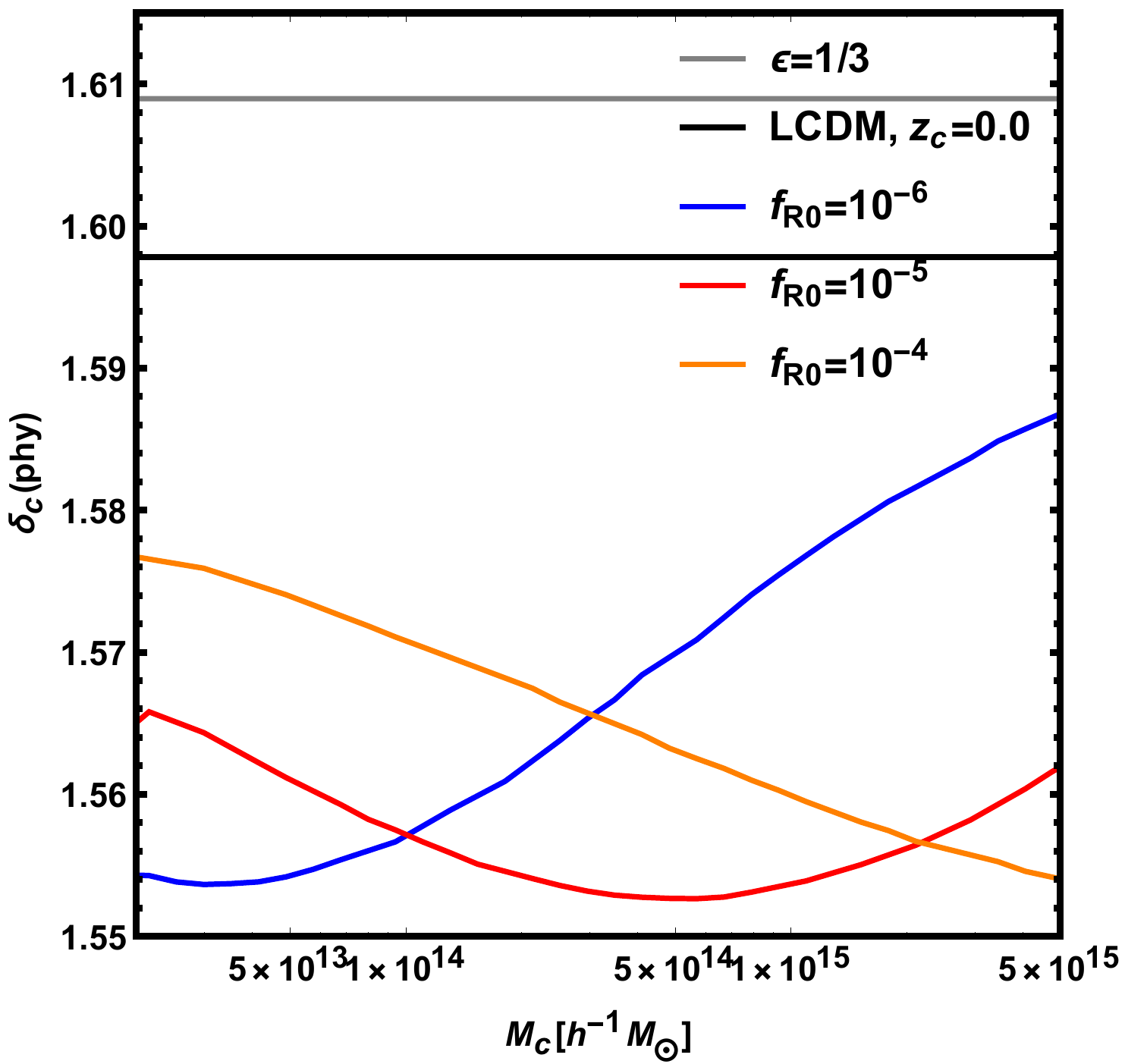}
\caption{\small{Top panels, from left to right: values of $\delta_c$ for the Tanh and Phy profiles, respectively, as a function of mass. Captions are identical to those of Fig. \ref{fig:compadcZdifRel}.
}}
\label{fig:compadcMdifRel}
\end{figure}

Fig. \ref{fig:compadcMdifRel} shows the dependence of $\delta_c$ with mass. The sensitivity to MG effects peaks at a level of $\sim2.7\%$, when $\delta_c \sim 1.55$. This takes place for masses $\sim 2 \times 10^{13} \, M_{\odot} \, h^{-1}$ if $|f_{R0}|=10^{-6}$,  for $M \sim 5 \times 10^{15} \, M_{\odot} \, h^{-1}$ if $|f_{R0}|=10^{-5}$, and for $M \gtrsim 5 \times 10^{15} \, M_{\odot} \, h^{-1}$ if $|f_{R0}|=10^{-4}$. 
These plots also show how varying the slope of initial profile impact $\delta_c$, with the largest effects appearing for smooth profiles ($s=0.8$).
This effect is also a consequence of the scale dependence of strength of the modifications of gravity and the concentration of the initial profile. Here also, as it happened with the time dependence, the SF limit is violated in both initial profiles.     

\section{Effects of \texorpdfstring{$f(R)$}{f(R)} on turnaround}
\label{sec:4}
\subsection{Time dependence}
\begin{figure}[ht]
\centering
\includegraphics[width=3.0in]{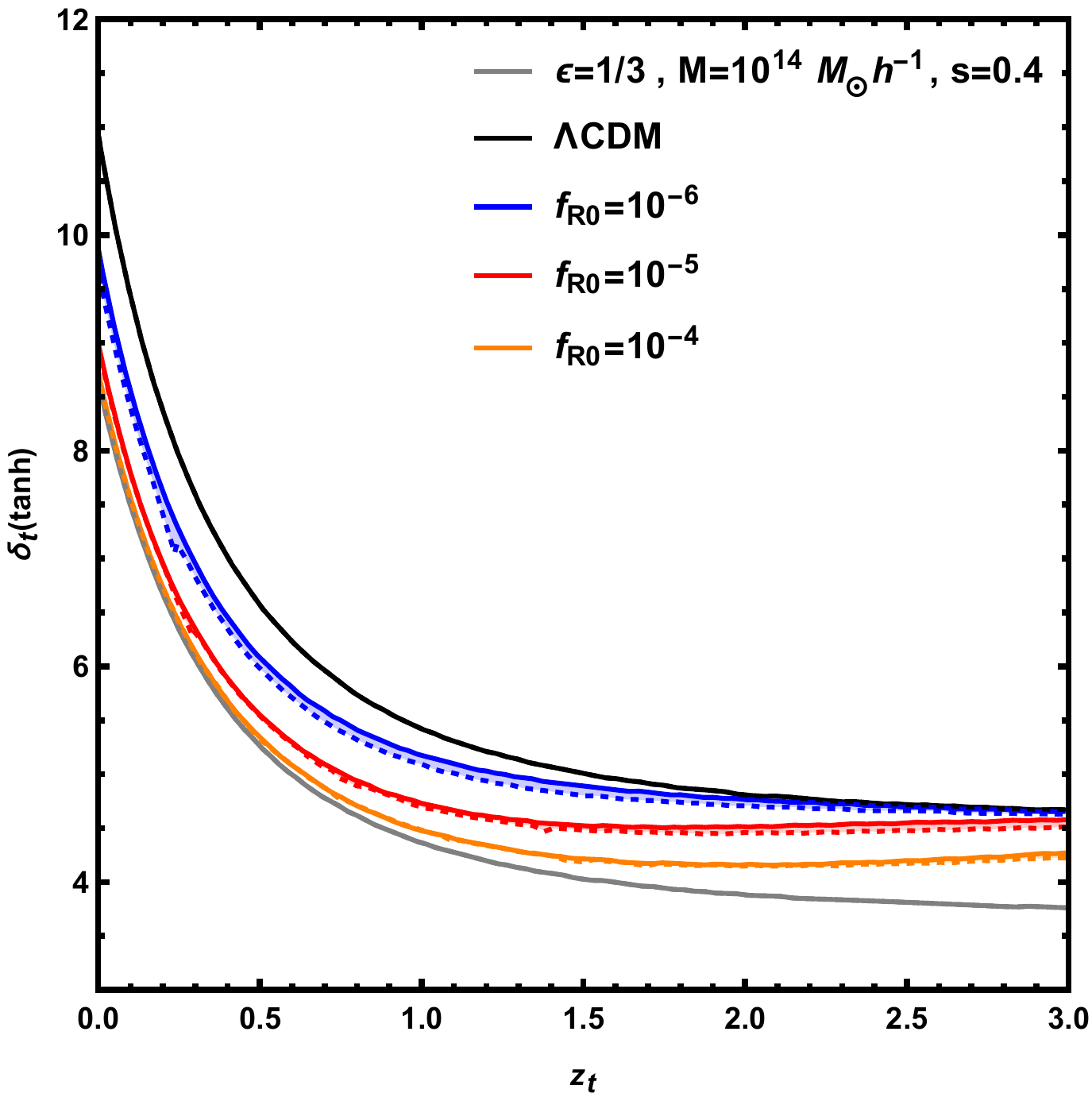}
\includegraphics[width=3.0in]{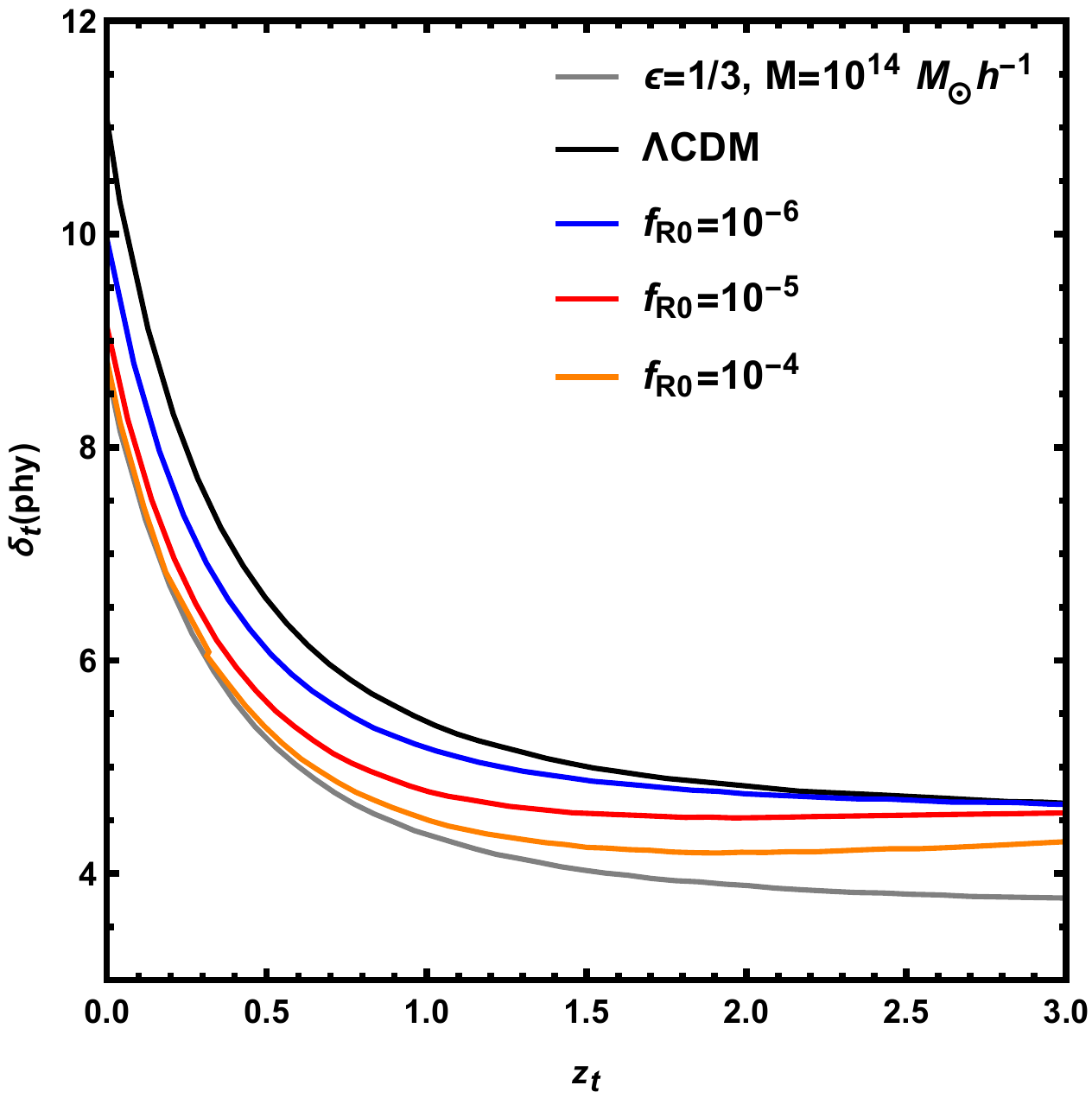}
\includegraphics[width=3.0in]{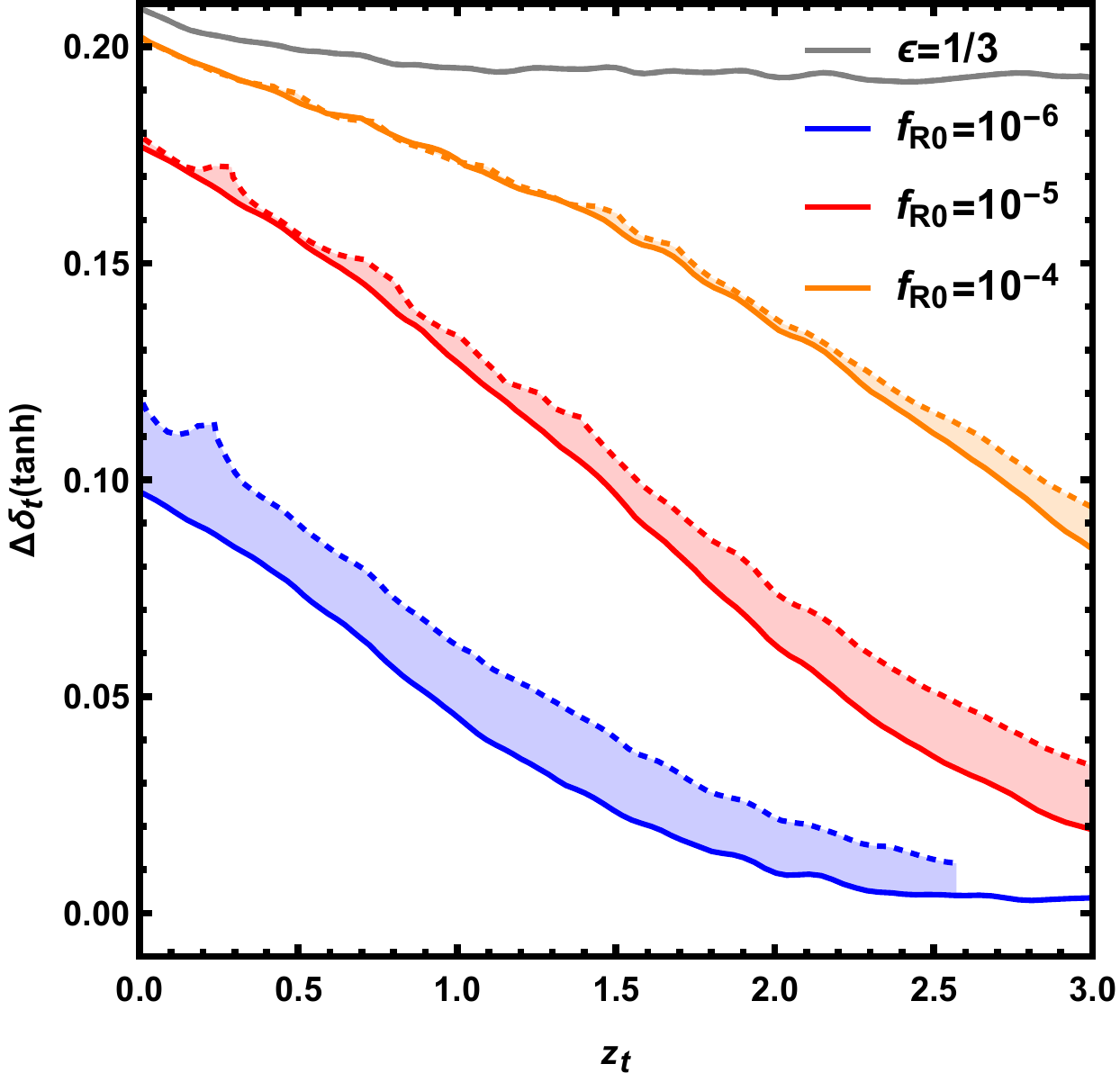}
\includegraphics[width=3.0in]{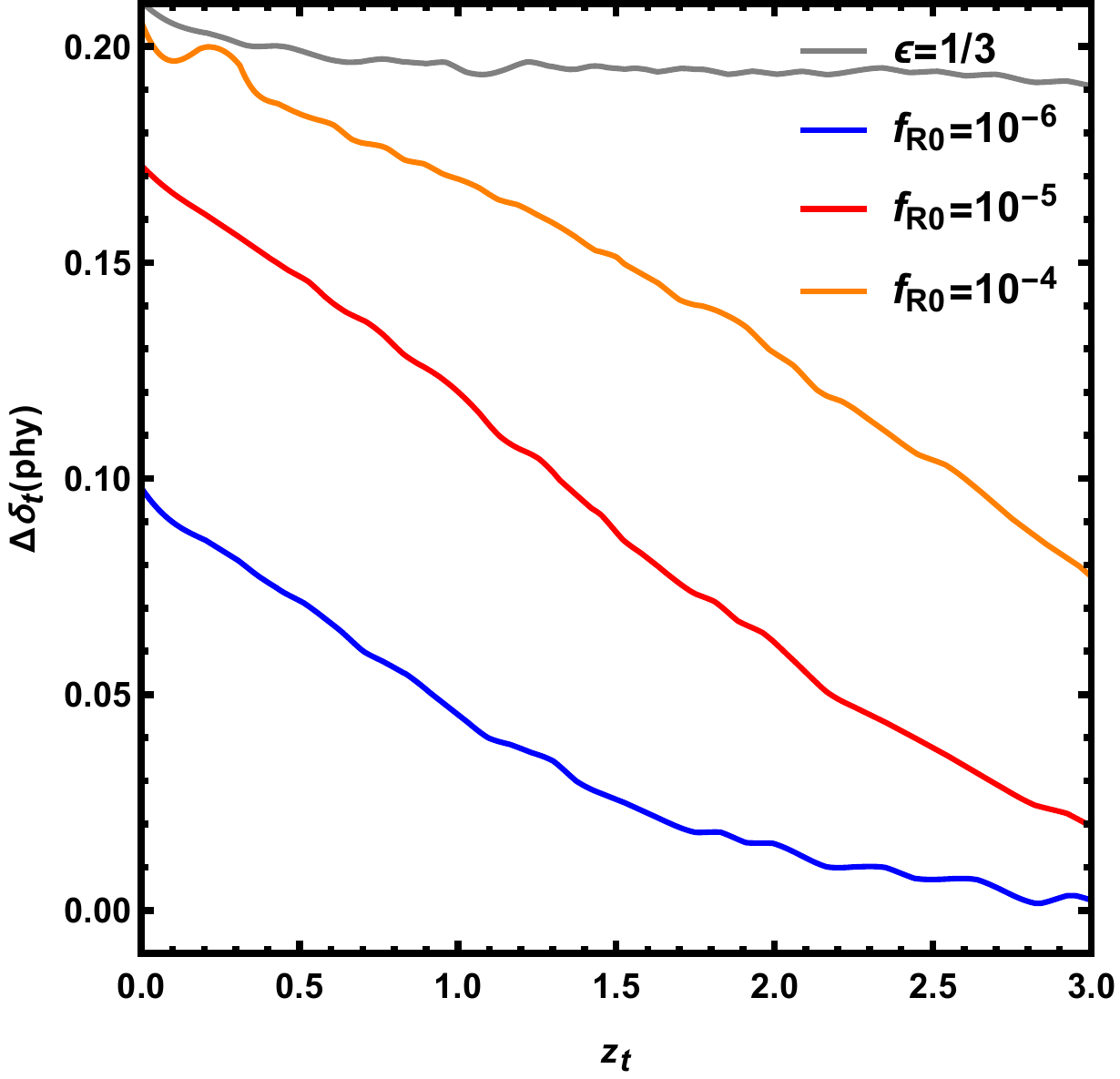}
\caption{Density contrast at turnaround, for the Tanh (left panels) and Phy (right panels) profiles, for $0<z<3$. 
Lower panels: relative differences between the values on the top panels with respect to $\Lambda$CDM. }
\label{fig:compadtZdifRel}
\end{figure}

The density contrast at the moment of turnaround, $\delta_t$, depicted in Fig. \ref{fig:compadtZdifRel}, is lower compared with the $\Lambda$CDM value by approximately $\sim 20\%$, $\sim 18\%$ and $\sim 10\%$ for $|f_{R0}|=10^{-4}$, $|f_{R0}|=10^{-5}$ and $|f_{R0}|=10^{-6}$, respectively, at $z \simeq 0$.
However, as opposed to the case for the critical density, $\delta_t$ does not surpass the SF limit ($\epsilon = 1/3$), and there is a remarkable consistency between the densities at turnaround amongst the different profiles.

Nevertheless, the effect of modified gravity on $\delta_t$ depends on the profile (and the slope in the case of the Tanh profile), and the deviation from $\Lambda$CDM is larger for higher values of $f_{R0}$. Also of importance is the fact that the relative differences increase at low redshifts.  These results show how the sensitivity of observables related to $\delta_t$ can be exploited in order to distinguish between MG and $\Lambda$CDM. This issue will be further explored in the next Section.
\subsection{Mass dependence}
The dependence of $\delta_t$ with mass is presented in Fig. \ref{fig:compadcMdifRel}. 
The sensitivity is higher in the low end of the mass range, where all values of $|f_{R0}|$ approach the LF limit, which is $\sim 20\%$ lower than for $\Lambda$CDM -- recall that for $\delta_c$ this difference was at most $\sim3\%$, with $s=0.8$. These plots also show that the slope of the initial profile has an effect, especially for higher masses, and the largest difference with respect to $\Lambda$CDM appears when $s=0.4$.
\begin{figure}[ht]
\centering
\includegraphics[width=3.0in]{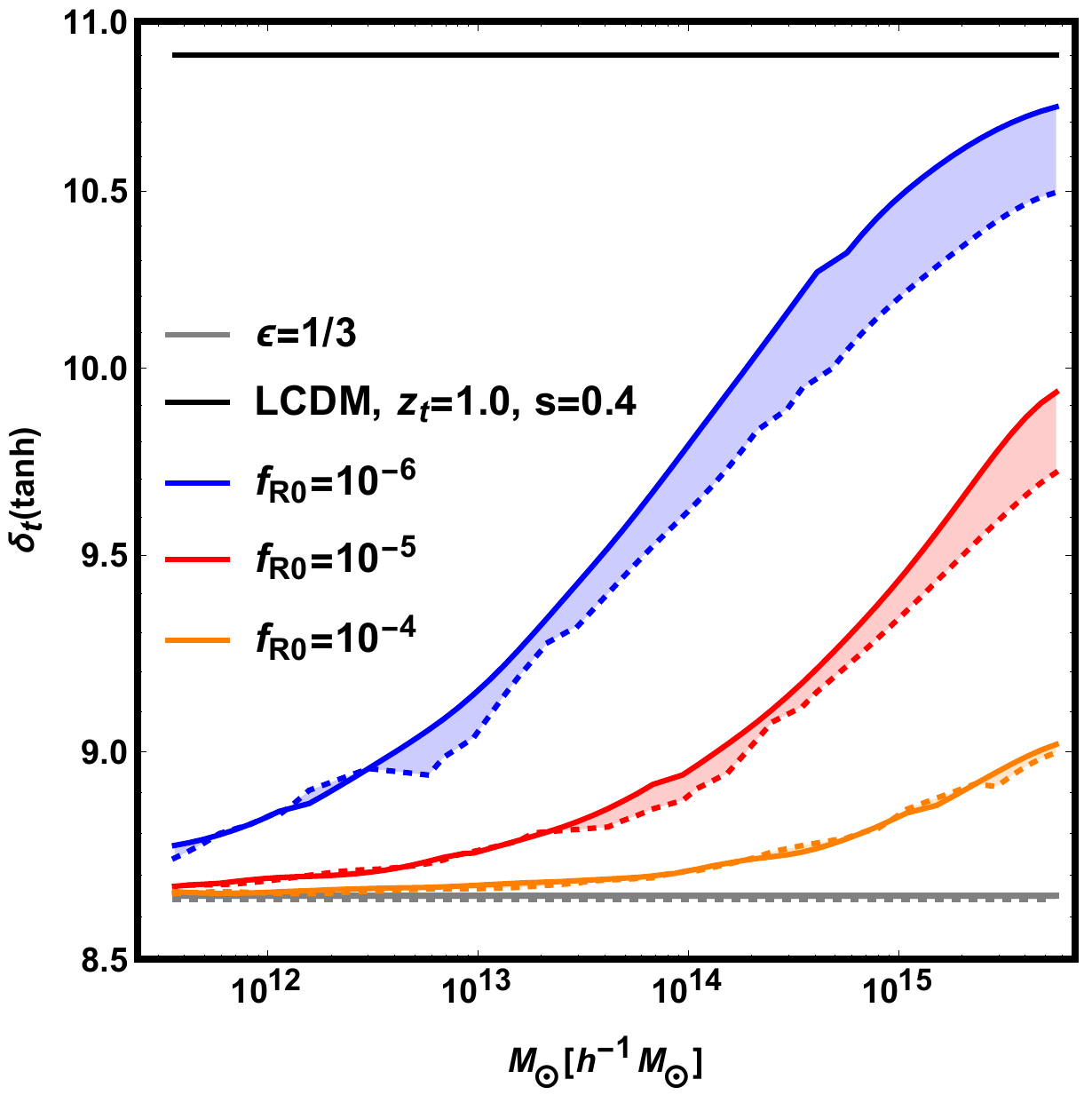}
\includegraphics[width=3.0in]{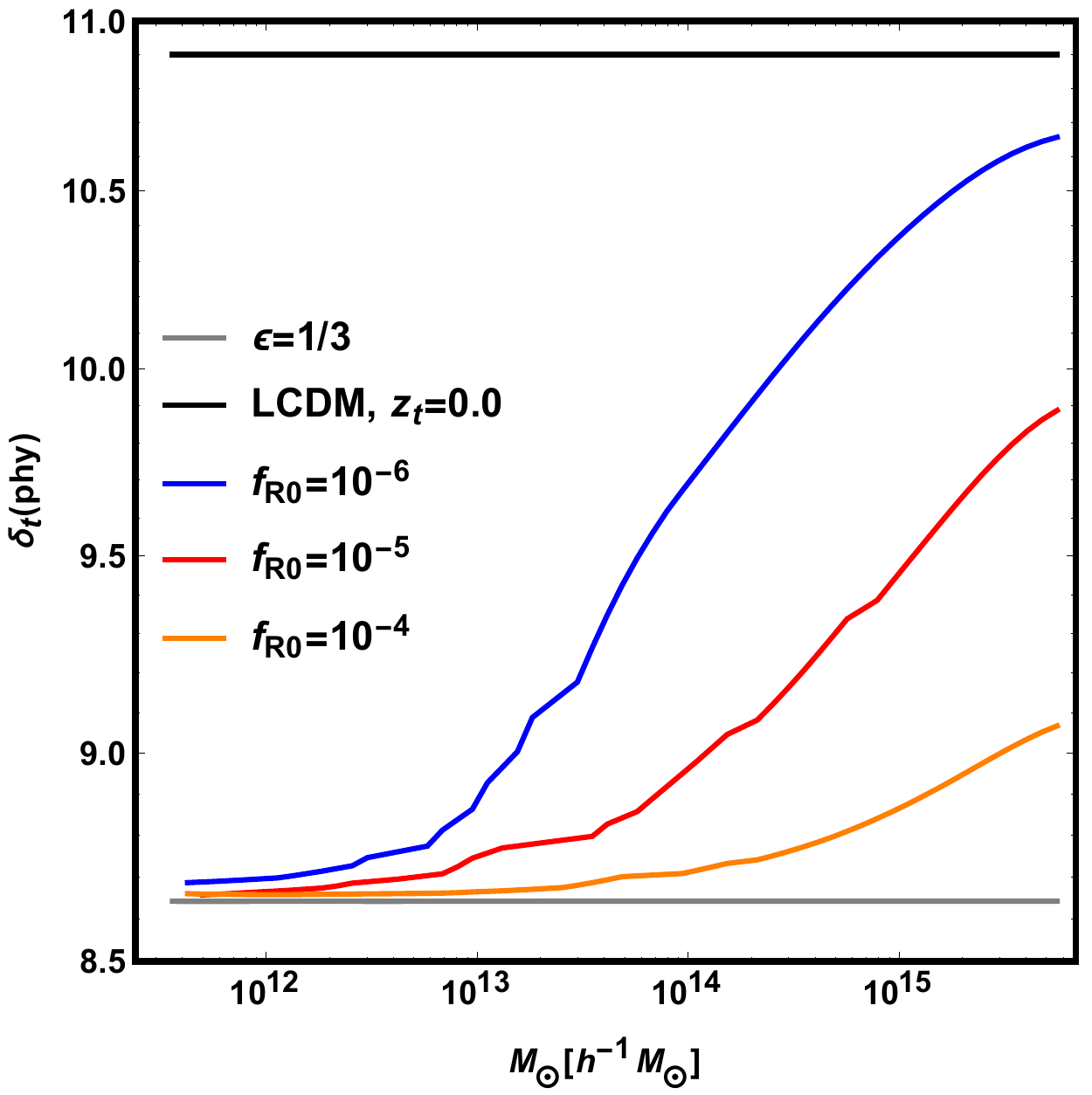}
\caption{Density contrast at turnaround as a function of mass, for the Tanh (left panel) and Phy (right panel) profiles.}
\label{fig:compadtMdifRel}
\end{figure}

\section{Turnaround radius in \texorpdfstring{$f(R)$}{f(R)} model}
\label{sec:5}
The size of a structure at the moment of turnaround, the so-called \textit{turnaround radius} ($R_t$), is a more direct observable compared with the density at turnaround, and its suitability as a cosmological test was studied by Refs. \cite{Pavlidou:2014aia, Pavlidou:2013zha, Tanoglidis:2014lea}. 
We have defined $R_t$ as the radius of the spherical shell around a structure that separates objects which fall towards the center from objects that follow the Hubble flow -- i.e., objects at this distance from the center should have null velocities. 
Theoretically, the competition between the gravitational attraction and the expansion due to the Hubble flow implies that there is a maximum size of $R_t$ for a structure of mass $M$, given by in $\Lambda$CDM  $R_{t,max}=\left(3GM/\Lambda c^2\right)^{1/3}$ \cite{Pavlidou:2013zha}. 
Therefore, it has been proposed that independent measurements of $R_t$ and $M$ could be used to test the value of $\Lambda$. Furthermore, estimates of $R_t$ for the galaxy group NGC5353/4 found hints of a possible violation of the maximum limit proposed in \cite{lee2015bound}.

These ideas have not yet been sufficiently explored in the context of MG. In \cite{faraoni2016turnaround}, $R_t$ was defined in terms of two perturbative  potentials, which are directly affected by modifications of gravity. On the other hand, \cite{bhattacharya2016large} have calculated an expression for $R_{t,max}$ in the context of the Galileon cubic model.
In $f(R)$ models, the effects on the maximum turnaround radius, $R_{t,max}$, was explored in \citep{2018arXiv180501233C}. On the other hand, they compute the radius of the last shell which collapses as a function of the mass of the structure, assuming only spherical symmetry.
We, in contrast, focus on the effect of MG in the shell which reaches the turnaround, as a function of time and of mass.

\subsection{\texorpdfstring{$R_t$}{Rt} and time dependence}
\begin{figure}[ht]
\centering
\includegraphics[width=3.0 in]{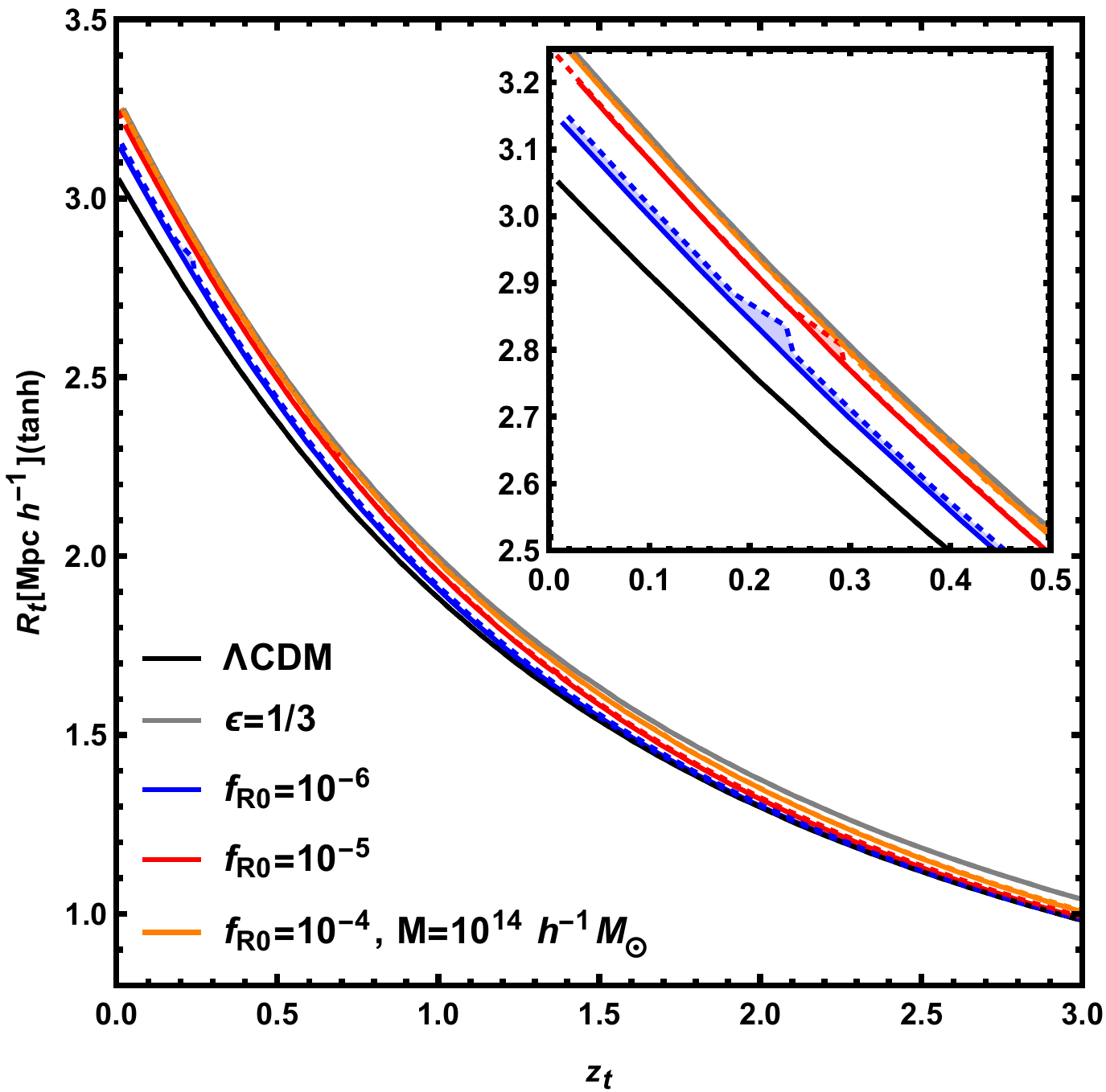}
\includegraphics[width=3.0 in]{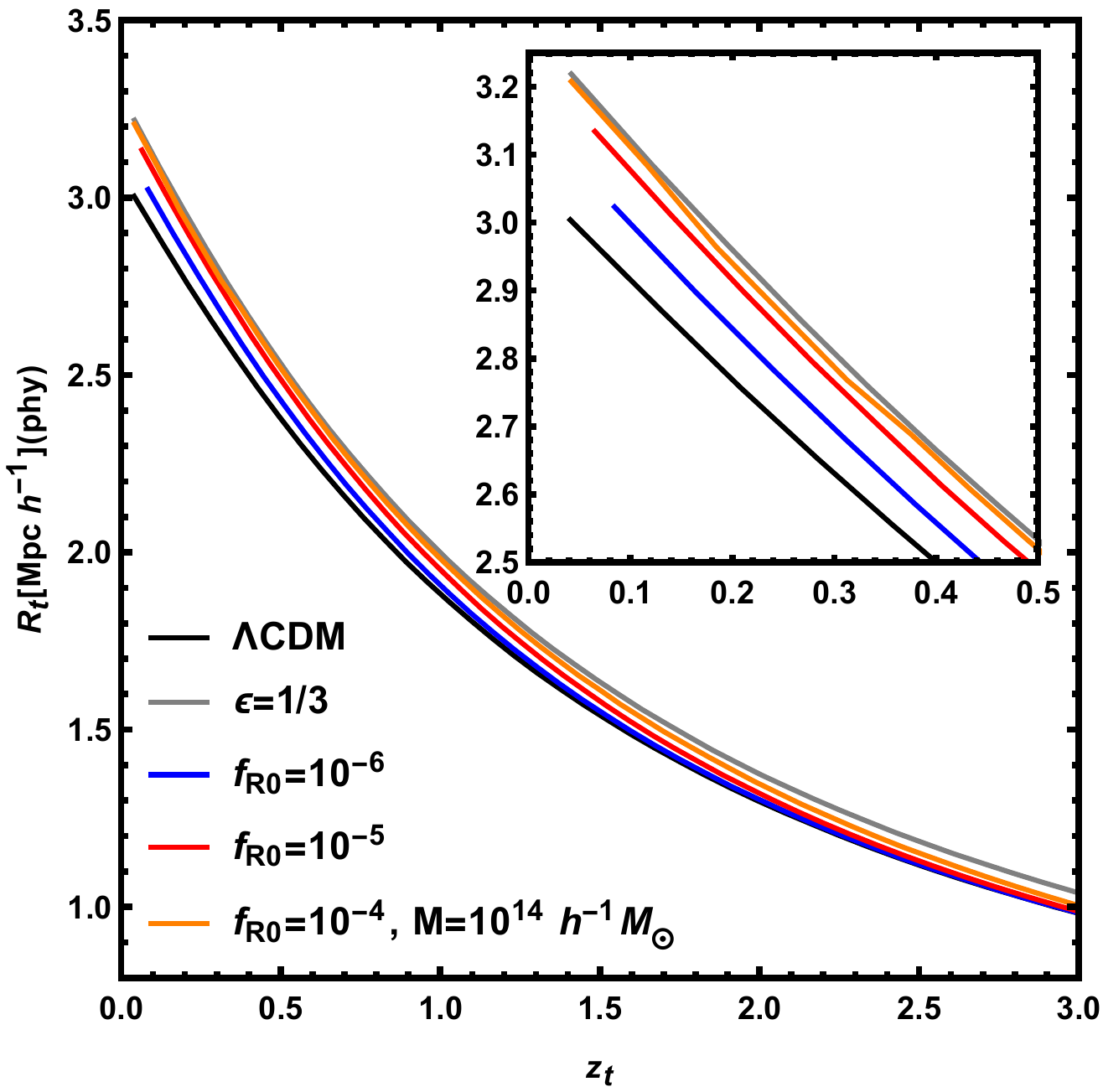}
\includegraphics[width=3.0 in]{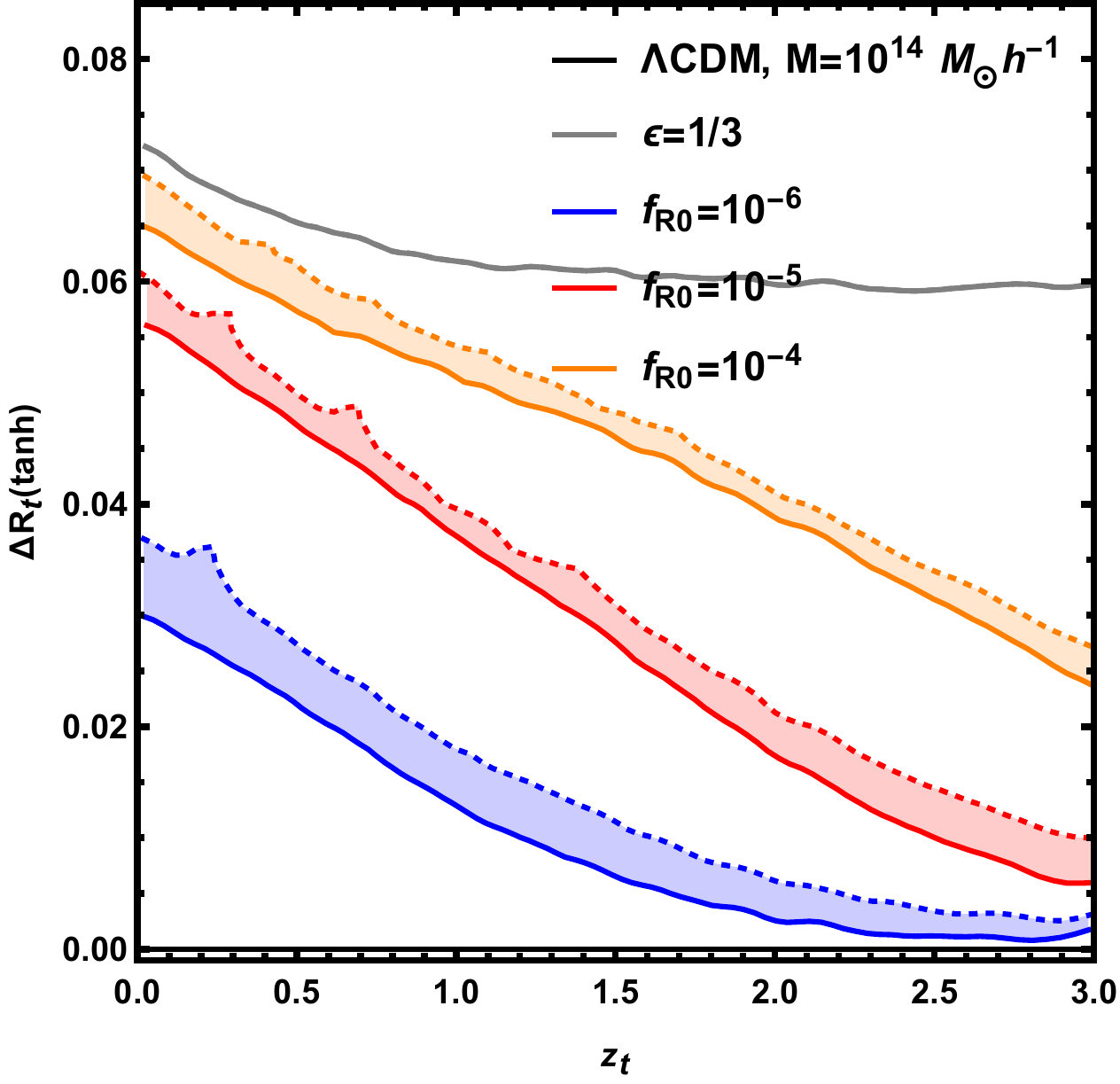}
\includegraphics[width=3.0 in]{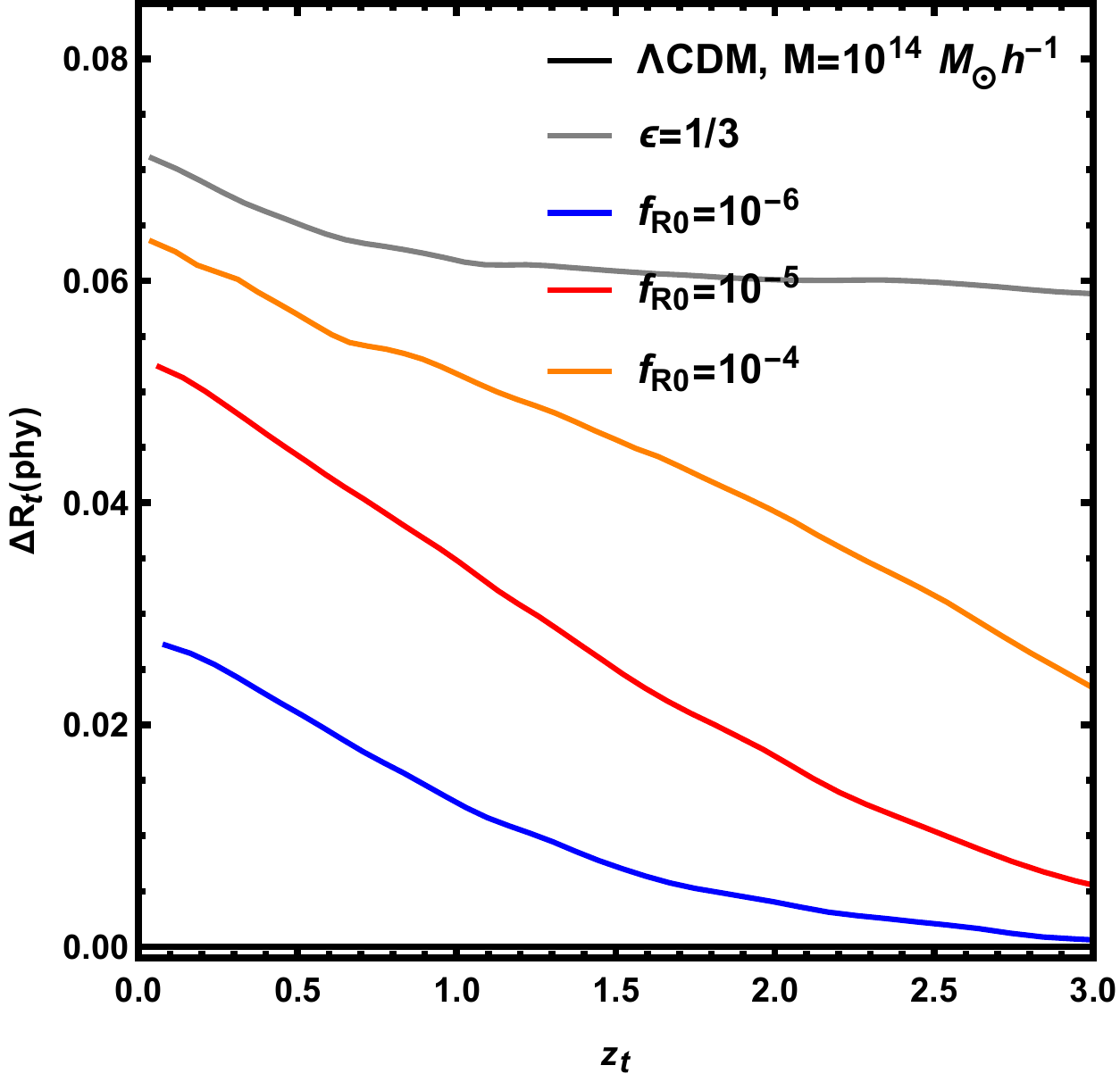}
\caption{Turnaround radius as a function of redshift, for structures with a mass of $M=10^{14}h^{-1}M_{\odot}$, for the Tanh (left) and Phy (right) profiles.}
\label{fig:rtztM14}
\end{figure}

We have seen that the properties of the turnaround and collapse of cosmic structures are sensitive to the modifications of gravity introduced in the $f(R)$ Hu-Sawicki model. 
In Fig. \ref{fig:rtztM14} we show the turnaround radius $R_t$ as a function of redshift in the several cases we have been considering: the SF limit ($\epsilon=1/3$), $|f_{R0}|=10^{-4}$, $10^{-5}$, $10^{-6}$ and $0$. As before, we also show results for the Tanh profile (left panels), with the two slopes $s=0.8$ (dotted) and $s=0.4$ (solid), as well as for the Phy profile (right panels).

Since the density of a spherically symmetric structure with an inner constant mass can be written as $\rho_m(a)=3M/4 \pi R^3(a)$, the physical turnaround radius also can be expressed as: 
\begin{equation}
R_t(a, M)=\left[\frac{3}{4 \pi \Omega_{m0} \rho_{c}[1+\delta_t(a, M)]}\right]^{1/3} a \, M^{1/3} \, ,
\label{eq:rtdeltat}
\end{equation}
where we used $\rho_m(a)=\bar{\rho}_m(a)[1+\delta(a, M)]$, with $\bar{\rho}(a)=\rho_{m0} \, a^{-3}$ and $\rho_c=2.77 \, h^{2} \, M_{\odot} \, {\rm Mpc}^{-3}$.
The turnaround radius $R_t(a, M)$ represents the radius of a shell whose  inner mass is $M$, and whose density contrast $\delta_t(a, M)$ is the density contrast when the first shell, near the center of the profile, reaches its maximum expansion. This moment corresponds to a scale factor $a_t$, so we can also write $R(a_t, M)\rightarrow R_t(a, M)$ -- and analogously, $\delta(a_t, M)\rightarrow \delta_t(a, M)$.

The left panel in Fig. \ref{fig:rtztM14} shows $R_t$ for structures with $M\sim10^{14}h^{-1}M_{\odot}$ in the redshift range $0<z<3$, for both initial profiles -- Tanh (left) and Phy (right).
As can be seen from the plots, the maximum size of these structures is always larger with respect to the SF/$\Lambda$CDM limit. This can be explained as a consequence of the enhancement of the gravitational strength, which leads to a larger reach of gravity. However, as happened for $\delta_t$, this change is confined between the two limiting cases, LF and SF.  
In the bottom panels we show the relative difference between the values of the turnaround radii  with and without the modifications on gravity. The largest relative difference occurs at $z=0$, where $\Delta R_t$ is of approximately $7.0\%$, $\sim 6.0\%$ and $\sim 3.0\%$ higher than the $\Lambda$CDM values for $|f_{R0}|=10^{-4}$, $|f_{R0}|=10^{-5}$ and $|f_{R0}|=10^{-6}$, respectively. Another interesting feature is that by decreasing the steepness of the profile ($s$), one also increases slightly $R_t$. 

Therefore, the main conclusion of this Section is that measurements of the turnaround radius that can reach ${\cal{O}}(5-10\%)$ precision at $z\sim 0$ can be used as a test of Hu-Sawicky $f(R)$ models.

\subsection{\texorpdfstring{$R_t$}{Rt} and mass dependence}
\begin{figure}[ht]
\centering
\includegraphics[width=3.0 in]{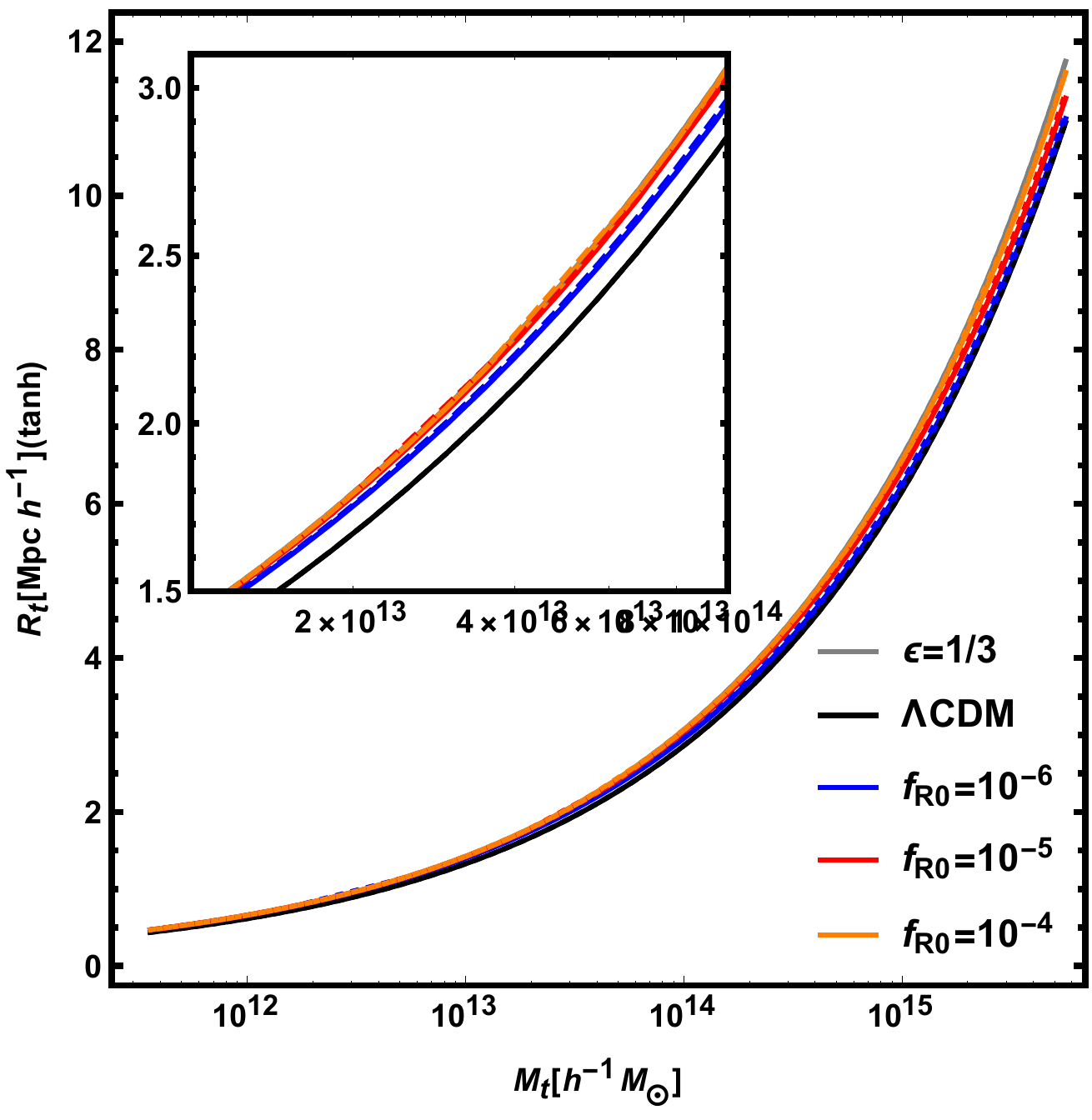}
\includegraphics[width=3.0 in]{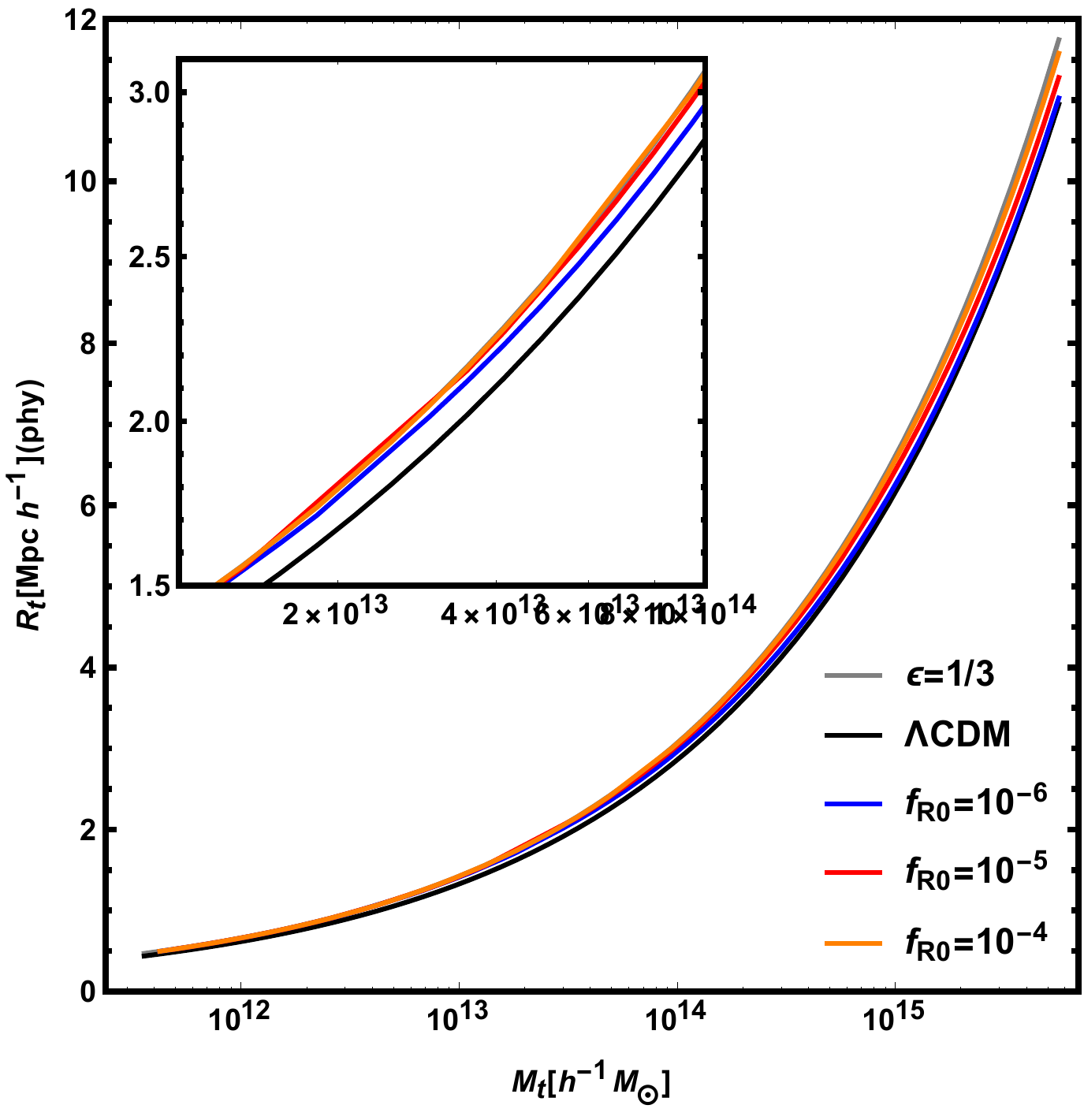}
\includegraphics[width=3.0 in]{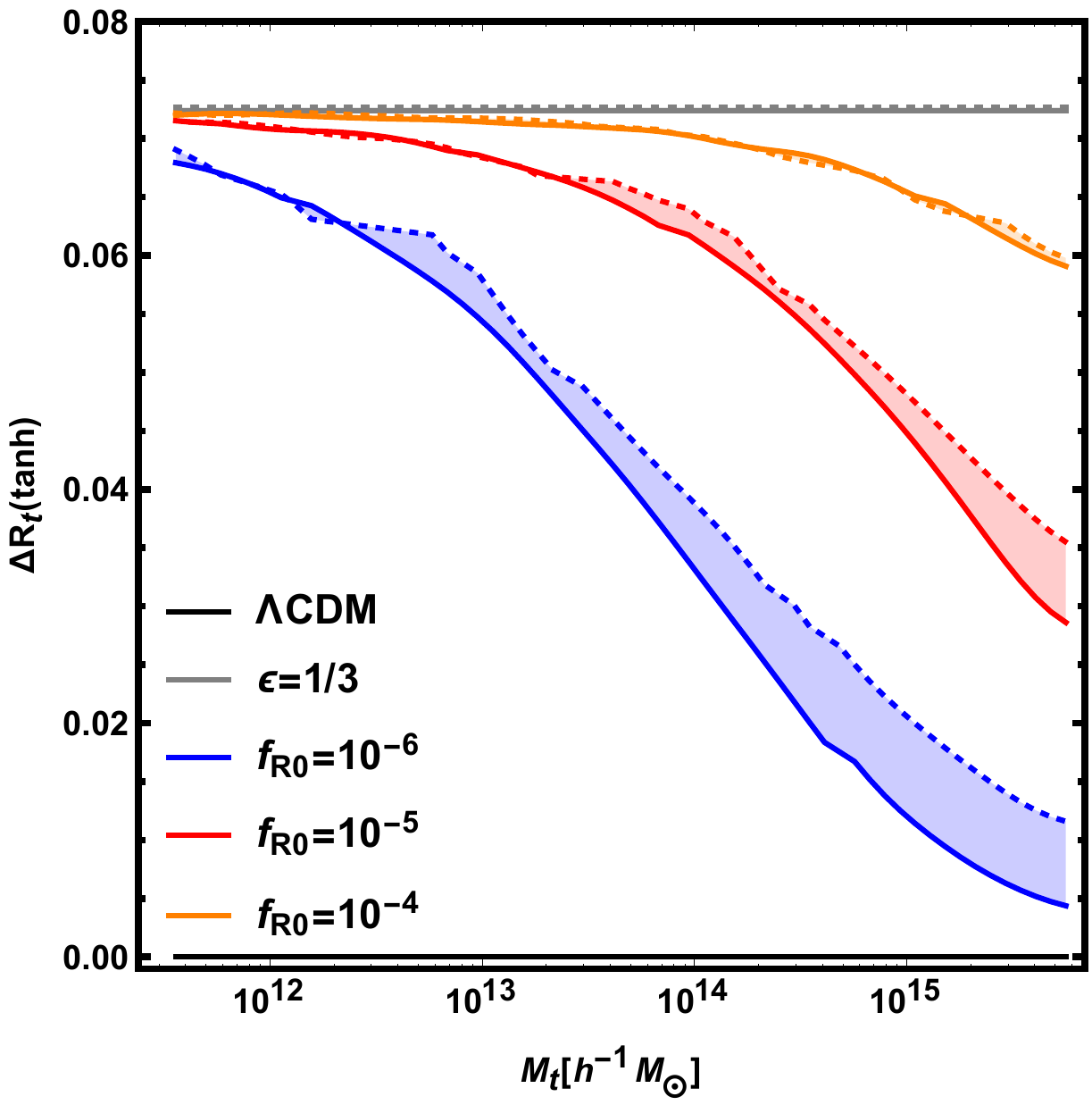}
\includegraphics[width=3.0 in]{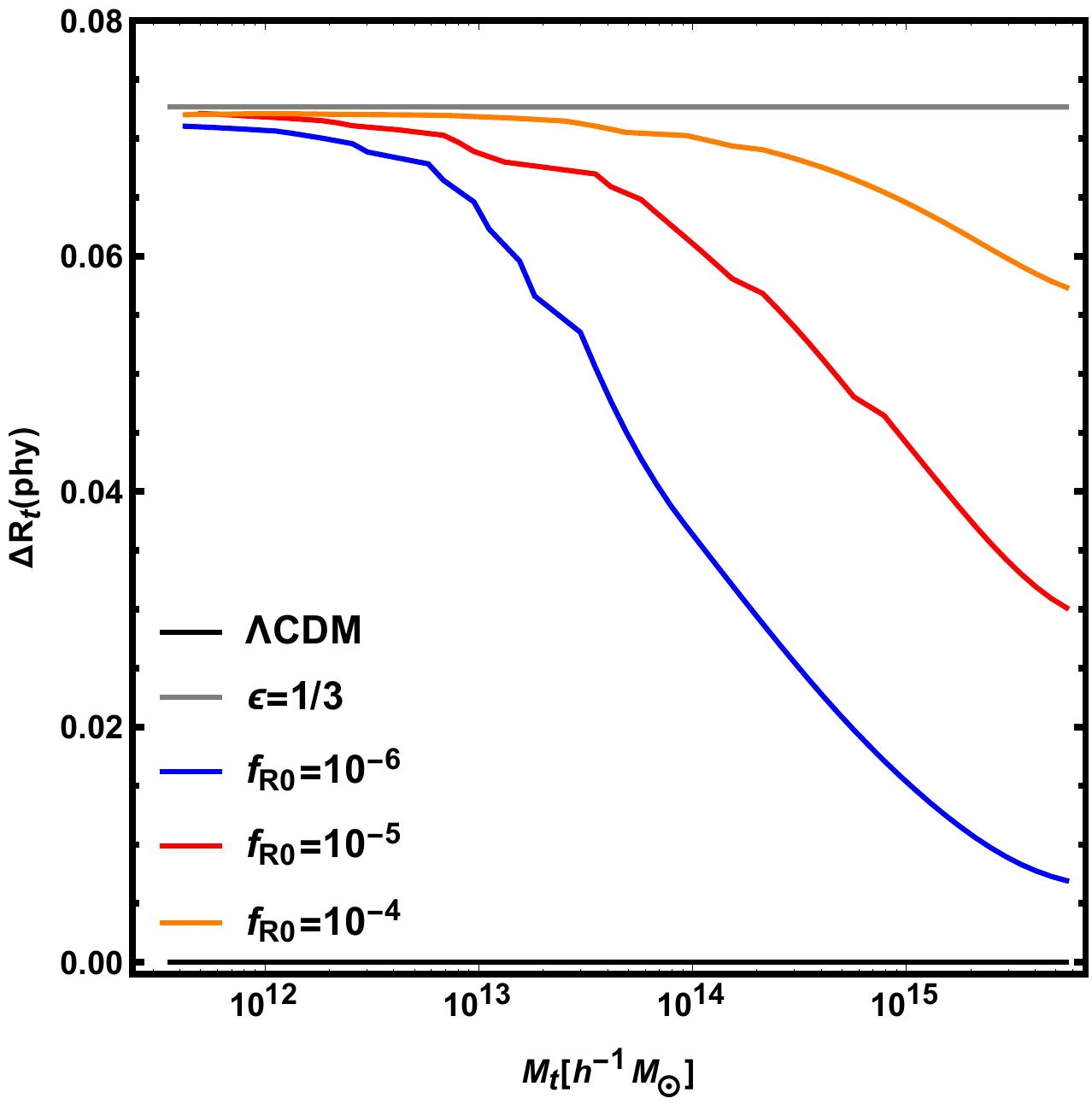}
\caption{Top panel: turnaround radius $R_t$ for masses in the range $10^{11} \,h^{-1}\, M_{\odot} < M < 10^{16} \,h^{-1}\, M_{\odot} \,$, for the Phy(left) and Tanh(right) profiles.  For structures with turnaround today.
The conventions for colors and lines are the same as previous figures.}
\label{fig:rtMtz0}
\end{figure}

A useful way of expressing the turnaround radius is through the relation between $R_t$ and the mass of the structure $M$ \cite{Pavlidou:2014aia, Pavlidou:2013zha}. In order to study the joint effects of the mass dependence, with varying strengths of the modifications of gravity, we show in Fig. \ref{fig:rtMtz0} the turnaround radius for the mass range $10^{11}h^{-1}M_\odot<M<10^{16}h^{-1}M_{\odot}$, for structures which reach their turnarounds at $z\simeq0$. As before, we consider the modified gravity parameter values $|f_{R0}|=10^{-4}$ (orange), $|f_{R0}|=10^{-5}$ (red), $|f_{R0}|=10^{-6}$ (blue), $\epsilon=0$ (black), the SF limit (gray), and $\epsilon=1/3$ (the LF limit), both for the Tanh profile (left) with slopes $s=0.8$ (dotted) and $s=0.4$ (solid), and for the Phy profile (right panels).

Fig. \ref{fig:rtMtz0} shows that the increase in $R_t$ is greater for smaller mass, in all MG models considered for this work. 
For smaller masses, corresponding to $M\simeq10^{12}h^{-1}\,M_{\odot}$, the deviation of $R_t$ reaches $\sim7\%$ with respect to the value in standard gravity, for all values of MG parameters, and in both profiles. As for the dependence on the slope of the initial profile, the turnaround radius is larger for increasing valus of $s$, but never exceeds $\sim 7\%$ of the standard value. Therefore, observations of structures with masses from $\sim 10^{12} h^{-1}M_\odot$ to $\sim 10^{14} h^{-1} M_\odot$, that achieve a precision of $4-8\%$ or better, can serve as a test of the type of MG model considered here.

\section{Conclusions}
\label{sec:conc}
The main focus of this work was to analyze whether the moment of turnaround could be used as way to test $f(R)$ models of modified gravity, namely the Hu-Sawicki model. We studied both the density contrast at turnaround, $\delta_t$, as well as the linear density contrast at the moment of collapse, $\delta_c$. Because of the scale-dependence of the modifications of gravity in $f(R)$ models, spherical collapse is a more complex mechanism compared with the top-hat density profile framework that describes the collapse of pressureless matter in $\Lambda$CDM. In contrast with the standard case, in MG the evolution depends both on the mass of the structure as well as on the density profile. We have modelled these features by considering a wide range of masses, as well as several different density profiles.

We have found that $\delta_t$ is affected by modifications of gravity, in such a way that its value decreases with respect to the $\Lambda$CDM values by up to $\sim 15\%$ if the modified gravity strength parameter $|f_{R0}|=10^{-5}$, for structures which reach their maximum sizes today. 

We also computed the turnaround radius in a variety of cases.
Even for the weakest version of MG parameters that we considered, where $|f_{R0}|=10^{-6}$, $R_t$ increases by $\sim 7\%$ for structures of $M \sim10^{12}\,h^{-1} \, M_{\odot}$. These results show the potential of observations that allow us to pinpoint regions that are experiencing turnaround today as a tool to study modifications of gravity on the scales of galaxy groups,  clusters and super-clusters.

Finally, we also point out that in this work the mass of the collapsing structure was measured in terms of the total matter inside the radius corresponding to the turnaround radius. However, due to the hierarchichal nature of structure formation, we expect turnaround to be happening continuously around structures which have already collapsed, on increasingly larger scales. 
Therefore, in realistic observations one would look for the turnaround radius in the outermost regions around collapsed (or virialized) halos. 
In that sense, the relevant mass to parametrize the turnaround radius would be the mass of the central halo, and not the total mass (which is very hard to measure due to the sparsity of the outermost regions). The relationship between these two masses, and how one can describe the turnaround in terms of a central/virial mass, will be the subject of a forthcoming paper.
\appendix
\section{Initial physical profile}
\label{append}
We follow the same approach of appendix C of \cite{PhysRevD.88.084015} to construct the initial physical profile, which was constructed using the peaks theory of Bardeen \textit{et al.}, 1986. \cite{bardeen1986statistics}. In this appendix we will review this construction of this profile for completeness.

We are interested in the mean density profile around some density peak of height $\nu = \delta _{0}/\sigma (R)$ of some Gaussian density field, that can be totally characterized by it matter power spectrum $P(k, R)$ smoothed in some scale $R$ by some window function $W(kR)$ (for these calculations was used a Gaussian window function).
    
Following these ideas, the $F(\nu,n_s,k,R)$ function, that is used to compute the initial density profile in ~\eqref{peaksshapeprimordial1}, is given by:
\begin{widetext}
\begin{multline}
F(\nu,n_s,k,R)=\\ 
   \left(\frac{e^{-\frac{1}{8} \left(\frac{n_s+3}{n_s+5}\right)^{3/2} \nu ^2} \left((12 n_s+60)
   e^{\frac{1}{8} \left(\frac{n_s+3}{n_s+5}\right)^{3/2} \nu ^2}+(0.632 n_s+13.52)
   n_s+44.6\right)}{(n_s+5)^2 \left(2 \sqrt{\frac{(0.25 n_s+0.75) \nu ^2+0.45
   n_s+8.25}{n_s+5}}\sqrt{\frac{n_s+3}{n_s+5}} \nu
   \right)}+\sqrt{\frac{n_s+3}{n_s+5}} \nu \right) \cdot \\  \phantom{\frac{\sqrt{\frac{n_s+3}{n_s+5}}}{ \Gamma \left(\frac{n_s+5}{2}\right)}}\cdot\left(\frac{\sqrt{\frac{n_s+3}{n_s+5}}}{ \nu \Gamma
   \left(\frac{n_s+5}{2}\right)}\right)\left(2 k^2
   R^2-n_s-3\right)+\frac{(n_s+3)  \left(-2 k^2 R^2+n_s+3\right)}{2 \Gamma
   \left(\frac{n_s+7}{2}\right)}+\frac{4 }{(n_s+5) \Gamma \left(\frac{n_s+3}{2}\right)}.
   \label{peaksshapeprimordial2}
\end{multline}
\end{widetext}

\section{Evolution of density profiles}
\label{append2}

\begin{figure}[ht]
\centering
\includegraphics[width=2.75in]{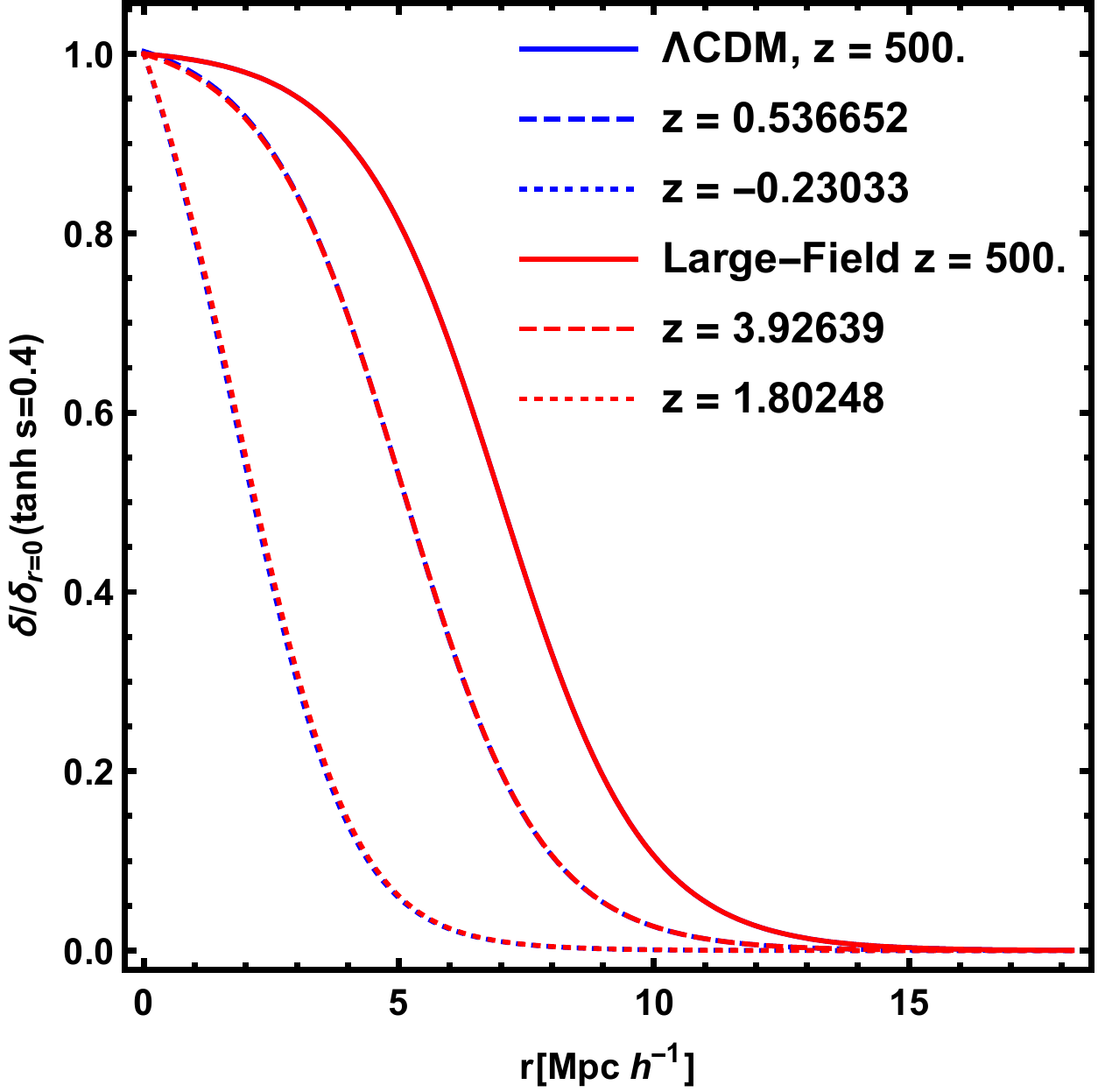}
\includegraphics[width=2.75in]{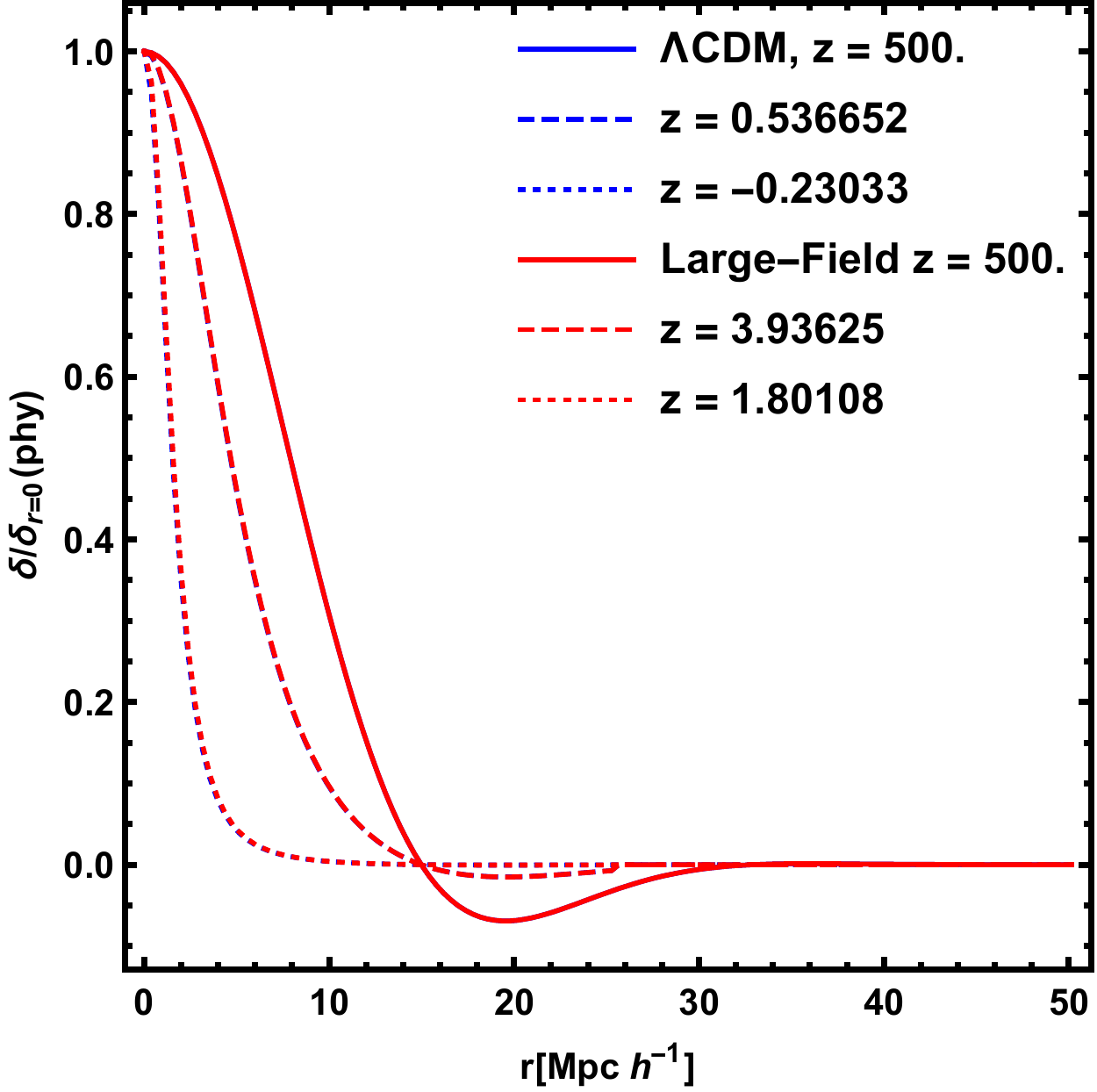}
\caption{Evolution of initial density profile, Tanh (left panel) and Phy (right panel), at initial (solid), intermediate (dashed) and collapse (dotted) moment, for GR (blue) and large-field gravity (red).}
\label{fig:profile_evol}
\end{figure}

In Fig. \ref{fig:profile_evol} we show the shape of the initial profile, for a mass of $ M=1.26 \times 10^{14}\,h^{-1} \, M_{\odot}$, as well as the profile shapes at two later instants, as the structure evolves, reaches an intermediate moment near the turnaround and approaches the moment of collapse. 
The evolved profile in the GR case (small-field limit of MG theories) is shown in blue, and MG in the large-field limit is shown in red, both for the Tanh (left) and Phy (right) profiles. 	
We have selected the profile at the moments when the central density has the same values for both limits, which shows explicitly that the profiles follow self-similar evolutions, but with different speeds in each limit. We also note, once again, that the differences in the redshifts between the two profiles are due to numerical difficulties to evolve Eq. \eqref{deltanonlinearx} in the case of the physical (Phy) profile.

\section{Turnaround radius and the maximum turnaround radius}
\label{append3}
\begin{figure}[ht]
 \centering
 \includegraphics[width=2.5in]{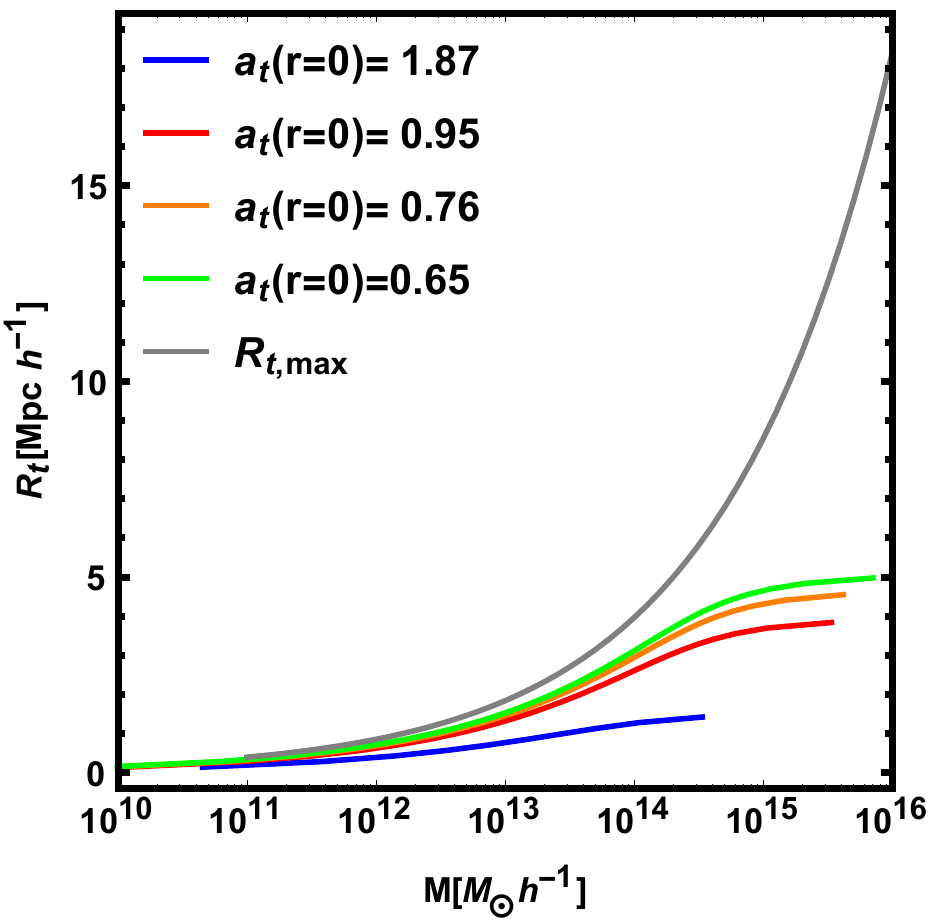}
 \caption{Turnaround radius of a structure for initial condition $\delta_0 = 0.005$ (blue), $\delta_0 = 0.0055$ (red), $\delta_0 = 0.006$ (orange) and $0.0065$ (green), in $\Lambda$CDM. With the maximum turnaround radius $R_{t,max}$ (gray).}
 \label{fig:shellMb}
 \end{figure}

With the goal of understanding the behavior of the shells of a given structure, we have evolved four profiles with different initial conditions, $\delta_0 = 0.005$ (blue), $\delta_0 = 0.0055$ (red), $\delta_0 = 0.006$ (orange) and $0.0065$ (green), in $\Lambda$CDM, and with $r_b=7$, where the shell near the center reaches turnaround at $a_t=1.87$, $a_t=0.95$, $a_t=0.76$ and $a_t=0.65$, respectively. 
Fig. \ref{fig:shellMb} shows the physical radius of each shell in the turnaround moment as a function of the mass inside this shell. We also plot the maximum turnaround radius obtained from the expression $R_{t,max}=\left(3GM/\Lambda c^2\right)^{1/3}$ \cite{Pavlidou:2013zha} (gray). We notice, in particular, that there is no violation of the maximum limit of the turnaround radius.

\acknowledgments
The authors would like to thank Marcos Lima for insightful comments on  structure formation in modified gravity models.
This work was supported by the {\it Programa Proqualis} of Instituto Federal de Educa\c{c}\~{a}o, Ci\^{e}ncia e Tecnologia do Maranh\~{a}o (RCCL) and by FAPESP (RV).
LRA and LSJ acknowledge support from FAPESP and CNPq.


%
%
%
%
%
%
%
%
\bibliographystyle{JHEP}
\bibliography{referencias}
\end{document}